\newif\ifarxiv
\arxivtrue 

\ifarxiv
	\documentclass[12pt]{article}
	\usepackage[margin=1in]{geometry}
	\usepackage[slantedGreek]{mathpazo}
	\usepackage[pdftex]{graphicx}
	\usepackage{amsfonts}
	\usepackage{amssymb}
	\usepackage{amsthm}
	\usepackage{graphicx,float}
 \usepackage{array, booktabs, makecell}
\usepackage{calc} 
 
\newtheorem{assump}{Assumption}[section]

\newtheorem{lemma}{Lemma}[section]

\usepackage{setspace}
\usepackage[dvipsnames]{xcolor}              
\usepackage{amsmath}%
\usepackage[
  colorlinks  = true,
  linkcolor   = {blue!50!black},
  urlcolor    = blue,
  citecolor   = {blue!50!black},
  linktocpage = true,
  ]{hyperref} 
\usepackage[round]{natbib}
    \usepackage{subfig}
\else
\fi

\newcommand\independent{\protect\mathpalette{\protect\independenT}{\perp}}
\def\independenT#1#2{\mathrel{\rlap{$#1#2$}\mkern2mu{#1#2}}}

\usepackage{color,soul,marginnote}

\ifarxiv
\title{Feature Selection for Personalized Policy Analysis}
\author{	
	\makebox[.4\linewidth]{Maria Nareklishvili  \footnote{
 Department of Economics, University of Oslo, Oslo, Norway and Booth School of Business, University of Chicago, Illinois, USA (Correspondence to: maria.nareklishvili@frisch.uio.no; nareklishvili@uchicago.edu)}}\\
    \and
	\makebox[.4\linewidth]{Nicholas Polson \footnote{
 Booth School of Business, University of Chicago, Chicago, Illinois, USA (Correspondence to:  ngp@chicagobooth.edu) }}\\
	\and 
	\makebox[.4\linewidth]{Vadim Sokolov  \footnote{
 Department of Computer Science, George Mason University, Fairfax, Virginia, USA. of Chicago, USA (Correspondence to:  vsokolov@gmu.edu) }}
}

\else
\fi

\graphicspath{{./fig/}}

\begin{document}
\ifarxiv
\maketitle

\begin{spacing}{1}
\begin{abstract}
In this paper, we propose Forest-PLS, a feature selection method for analyzing policy effect heterogeneity in a more flexible and comprehensive manner than is typically available with conventional methods. In particular, our method is able to capture policy effect heterogeneity both within and across subgroups of the population defined by observable characteristics. To achieve this, we employ partial least squares to identify target components of the population and causal forests to estimate personalized policy effects across these components. We show that the method is consistent and leads to asymptotically normally distributed policy effects. To demonstrate the efficacy of our approach, we apply it to the data from the Pennsylvania Reemployment Bonus Experiments, which were conducted in 1988-1989. The analysis reveals that financial incentives can motivate some young non-white individuals to enter the labor market. However, these incentives may also provide a temporary financial cushion for others, dissuading them from actively seeking employment. Our findings highlight the need for targeted, personalized measures for young non-white male participants.
\end{abstract}
\end{spacing}

\

\else
\fi
\section{Introduction}
Randomized control trials play an important role for treatment, policy, or a program effect analysis in economics, statistics, medicine, and other fields (\citealp{banerjee2009experimental, bertrand2017field, chernozhukov2018generic}).  To design and implement effective interventions, policymakers and researchers are often interested in partitions of the population that are particularly susceptible to a new program or a policy. Identifying such subgroups can be challenging, especially when there are a large number of observable characteristics that can influence the outcome. In such cases, conventional estimation methods, such as ordinary least squares,  may produce inefficient estimates. This is because these methods  often struggle to identify relevant variables in a sparse feature space \citep{johnstone2009statistical, belloni2013least}.

We propose Forest-PLS, a data-driven approach for selecting the  \textit{target components} of the population for personalized policy analysis. In our approach, the target components represent linear combinations of the explanatory variables. These components are characterized by large weights on the variables that are most strongly associated with the outcome, or carry a significant amount of information for explaining policy effect heterogeneity.  As a second step, we identify and estimate personalized policy effects across the chosen components. By focusing on these key components, rather than considering the full set of characteristics,  policymakers can design targeted interventions tailored to the most diversified segments of the population.

The procedure is a combination of two distinct methods, the partial least squares  \citep{geladi1986partial, vinzi2010handbook} and causal forest \citep{athey2016recursive, wager2018estimation} algorithms. The partial least squares method is used to detect the policy-relevant components in the first step. These components represent a reduced explanatory variable space. The reduced space reflects the highest variation in explanatory features and the most relevant information for predicting the outcome. These components are continuous rather than discrete clusters of the data space, allowing us to analyze policy effects across a full spectrum of the population segments. In the second step, we use the causal forest algorithm to identify different quantiles of policy effects within each component value. This allows us to capture heterogeneity in the policy effects at a finer granularity.

The primary contribution of this paper is to advance our understanding of the distribution of policy effects from a theoretical and empirical perspective. The theoretical component of the article demonstrates that our approach is consistent and leads to asymptotically normally distributed policy effects. Our framework and findings extend beyond a single coefficient of interest \citep{wager2018estimation} to multiple (plausibly) correlated policy effects. The empirical contribution of the paper is to identify and analyze two types of heterogeneity of the policy effects: within-group heterogeneity and between-group heterogeneity. The proposed method allows us to estimate quantiles of policy effects within and across the values of target components. Analyzing individual explanatory variables separately \citep{meinshausen2006quantile} can be challenging in high-dimensional data. Our approach allows us to focus on aggregate aspects of these characteristics without losing the economic interpretation of the resulting subgroups. 

Our framework is closely related to the papers dedicated to the estimation of personalized treatment effects \citep{athey2015machine, athey2016recursive, wager2018estimation, chernozhukov2018generic, chernozhukov2018sorted, kunzel2019metalearners, hahn2020bayesian, nie2021quasi, xiong2021federated}, and 
feature selection for the inference on treatment effects \citep{belloni2012sparse, belloni2014inference, chernozhukov2015valid, chernozhukov2015post, urminsky2016using, banerjee2021selecting}.  Previous work for personalized treatment effect analysis considers a single source of heterogeneity, such as quantile treatment effects, or treatment effects across the original set of covariates (features). We unify the feature selection methods with personalised policy analysis. This allows us to investigate a full density of policy effects within and across a pooled variable space (target components).

Other related methods are proposed by \cite{ hahn2002capturing, chun2010sparse, mehmood2012review, mehmood2020comparison, polson2021deep, nareklishvili2022deep, dixon2022deep} that use the partial least squares algorithm for dimension reduction. These methods suggest that partial least squares as a precursor to a more general framework of deep learning and instrumental variables can increase efficiency. By comparison, our study shows that the method can pool statistically and economically significant variables for policy effect heterogeneity. Additionally, \cite{nekipelov2018moment, li2020asymptotic, nareklishvili2022adaptive} investigate large sample properties for random forests under multiple outcomes, coefficients or network effects. We show that the theoretical properties hold even after the feature selection procedure by partial least squares. \cite{chernozhukov2018generic} and \cite{jacob2019group} propose group average 
One notable advantage of the Forest-PLS method is its ability to estimate heterogeneous group-average policy effects (GATE) in situations where the groups involved are both unordered and continuous. This characteristic distinguishes Forest-PLS from the approaches presented by  \cite{chernozhukov2018generic} and  \cite{jacob2019group}, which primarily focus on estimating GATE for ordered. It is important to note that the Bayesian approach is an alternative to policy effect heterogeneity. \cite{ansari2000hierarchical, taddy2016nonparametric, santos2018tree, hahn2020bayesian, woody2020estimating, starling2021targeted, krantsevich2022stochastic, he2023stochastic} formulate Bayesian Additive Regression Trees (BART) for heterogeneous treatment or policy effect analysis. The advantage of the approach lies in the regularization effect through predetermined priors of the tree parameters. Our work can be extended to accommodate Bayesian priors.

We design various simulated experiments to unveil the inherent predictive advantages of the approach when compared to the traditional benchmark algorithms. The results reveal two notable advantages of Forest-PLS over the causal forest algorithm. First, Forest-PLS exhibits remarkable resilience in recovering the true density of policy effects, even when confronted with a limited number of observations. This characteristic renders the algorithm highly robust and consistent across varying sample sizes. Second, Forest-PLS effectively mitigates the influence of redundant variables and noise present in the experimental setup, enabling accurate estimation of the variance of policy effects.

This article examines the impact of financial incentives on unemployment duration based on data from the Pennsylvania "Reemployment Bonus" Demonstration, a randomized control trial conducted in 1988-1989. The analysis reveals significant variation in the policy effects both within and across different subgroups of the population. Specifically, the results show that the effects of the policy are more significantly dispersed for young, non-white male claimants who joined the experiment early on, compared to middle-age and older female participants with a high number of dependents. The difference between the 97.5th and 2.5th percentiles of policy effects is 92.8\% for the first vigintile of the target component, and decreases to 22.1\% for the final vigintile of the component. These findings highlight the need for targeted, personalized measures for specific subgroups, such as young non-white male participants.

\section{The Forest-PLS Framework}
Consider the outcome $y_i \in \mathbb{R}$ (e.g., unemployment duration) for a subject $i = 1, \dots, N$. Each subject is characterized with an observable vector of features $\mathbf{X}_{i} \in \mathbb{R}^p$ (e.g., age, gender, occupation). We assume that $p = N^\psi$, with $\psi < 1$.  A policy is denoted by $P_i \in \{0, 1\}$, and we let $y_i(1)$ and $y_i(0)$ denote potential outcomes with and without the policy, respectively. We assume, the unconfoundedness holds:

\begin{assump}[Unconfoundedness]\label{ass_unc} The policy is independent of the potential outcomes\footnote{This assumption is stronger than the "no unmeasured confounders" assumption proposed by \cite{rosenbaum1983central}. Following \cite{rosenbaum1983central}, Assumption \ref{ass_unc} implies that if $P_i$ is unconfounded given $\mathbf{X}_i$, then $P_i$ is unconfounded given $g(\mathbf{X}_i)$, where $g$ is an affine transformation of $\mathbf{X}_i$. Intuitively, the transformation does not introduce any additional information beyond what is already known by $\mathbf{X}_i$. }:
\begin{align*}
    y_i(1), y_i(0) \independent P_i|\mathbf{X}_i. 
\end{align*}
\end{assump}

A policymaker wishes to identify the dimensions of the feature space that contain the most relevant information about the policy effects. To this end, we consider a mapping $f: \mathbf{X}_{i} \mapsto \mathbf{C}_{i} \in \mathbb{R}^q$ that maps the original features to a set of \textit{target components} $\mathbf{C}_i$. In other words, $\mathbf{C}_i$ is a collection of $q$-dimensional linear combinations of the features $x_{ij}$ for $j = 1, \dots, p$ (with $q \leq  p$). This transformation allows us to focus on a smaller, more interpretable set of features while preserving the information about the policy effects.
 
The coefficient of interest is the effect of $P_i$ on the outcome:
\begin{align*}
    \theta_{i}= y_{i}(1) - y_{i}(0),
\end{align*}

Personalized policy effects are not directly observable, as an individual is only exposed to one policy state (either with or without the policy). Therefore, we typically consider the expectations of the potential outcomes:

\begin{align*}
    \theta(\mathbf{C}_i) = \mathbb{E}(\theta_i|\mathbf{C}_i) + \varepsilon_i, \\
    \mathbb{E}(\theta_i|\mathbf{C}_i) = \mathbb{E}\big[y_{i}(1) - y_{i}(0)\big|\mathbf{C}_i].
\end{align*}
Due to Assumption \ref{ass_unc}, the policy effect of interest is given as 
\begin{align}\label{eq_ate}
   \tau(\mathbf{C}_i) = \mathbb{E}\big(y_i(1) -  y_i(0)|\mathbf{C}_i\big) =  \mathbb{E}\big(y_i|P_i = 1,\mathbf{C}_i\big)  -   \mathbb{E}\big(y_i|P_i = 0,\mathbf{C}_i\big),
\end{align}
where $\mathbb{E}\big(y_i|P_i = d,\mathbf{C}_i\big)$ for $d \in \{1, 0\}$ denotes the observed expected outcomes with and without the policy, respectively. 

To estimate the average policy effect in \eqref{eq_ate}, we need to determine the optimal target components $\mathbf{C}_i$ and use a method that estimates group-level policy effects, conditional on the chosen components.

\subsection{Identification of Target Components}
We seek to identify the linear combinations of features, also known as target components/scores/factors, $\mathbf{C} = [\mathbf{c}_1, \mathbf{c}_2, \dots, \mathbf{c}_q$] (i.e., $\mathbf{c}_1 = \mathbf{X}\mathbf{w}_1$) that explain the highest variation in covariates $\mathbf{X}$, as well as the outcome $\mathbf{y}$ \citep{tobias1995introduction, abdi2003partial}. $\mathbf{X}$ and $\mathbf{y}$ can be decomposed as:

\begin{align}
   \underset{(N \times p)}{\mathbf{X}} = 
  \underset{(N \times q)}{\mathbf{C}} \  \underset{(q \times p)}{\mathbf{V}^T} + \underset{(N \times p)}{\mathbf{E}}, \\
  \underset{(N \times 1)}{\mathbf{y}} = 
  \underset{(N \times q)}{\mathbf{C}} \  \underset{(q \times 1)}{\mathbf{b}} + \underset{(N \times 1)}{\mathbf{e}},
\end{align}
where $\mathbf{X} = [\mathbf{x}_1, \mathbf{x}_2, \dots, \mathbf{x}_p]$ is the matrix of covariates, $\mathbf{V}^T$ is the matrix of loadings (weights),  and $\mathbf{E}$ is the matrix of errors for the covariates. $\mathbf{y}$ denotes the outcome as before, $\mathbf{b}$ is the vector of coefficients (the influence of components on the outcome), and $\mathbf{e}$ is the vector of errors for the response. 

We use the iterative procedure to obtain the target components (aka \textit{partial least squares}). Consider, the weight $\tilde{\mathbf{w}}_1 = \big( cov(\mathbf{x}_1, \mathbf{y}), cov(\mathbf{x}_2, \mathbf{y}), \dots, cov(\mathbf{x}_p, \mathbf{y}) \big) = \big(\tilde{w}_{11}, \tilde{w}_{21}, \dots, \tilde{w}_{p1} \big)$. We normalize it to get a unit vector: 
\begin{align*}
    \mathbf{w}_1 = \frac{\tilde{\mathbf{w}}_1}{||\tilde{\mathbf{w}}_1||},
\end{align*}
where $||\tilde{\mathbf{w}}_1||$ denotes the Euclidean norm.  We use these weights to compute the first principal component:

\begin{align}\label{eq_first_component}
    \mathbf{c}_1 = w_{11}\mathbf{x}_1 + w_{21}\mathbf{x}_2 +  \dots + w_{p1}\mathbf{x}_p = \mathbf{X}\mathbf{w}_1 = \frac{\mathbf{X}\mathbf{w}_1}{\mathbf{w}_1^T\mathbf{w}_1}.
\end{align}
The last equality in \eqref{eq_first_component} follows by the fact that the weights are unit vectors. A linear regression of a $j-$th covariate on the first component yields a loading. The vector of loadings associated with the first component is given by regressing the covariates on it:
\begin{align}
    \mathbf{v}_1 = \frac{\mathbf{X}^T\mathbf{c}_1}{\mathbf{c}_1^T \mathbf{c}_1}.
\end{align}

Similarly, the first coefficient $b_1$ is obtained by regressing the outcome on the first component:
\begin{align}
     b_1 = \frac{\mathbf{y}^T\mathbf{c}_1}{\mathbf{c}_1^T \mathbf{c}_1}.
\end{align}

The next step is to obtain the approximation of the covariate matrix and the outcome, and predict residuals:
\begin{align}
    \mathbf{X}_1 = \mathbf{X} - \hat{\mathbf{X}},
    \mathbf{y}_1 = \mathbf{y} - \hat{\mathbf{y}},
\end{align}
where $ \hat{\mathbf{X}} = \mathbf{c}_1\mathbf{v}_1^T$ and $\hat{\mathbf{y}} = b_1\mathbf{c}_1$. We obtain the subsequent components by repeating the described procedure for the first, second, and higher order residuals of the covariate matrix $\mathbf{X}_1, \mathbf{X}_2 \dots$, and the outcome $\mathbf{y}_1, \mathbf{y}_2 \dots$, respectively.

A desirable property of the procedure is that the coefficients have a closed-form solution. The estimator of these coefficients is given as (\citealp{helland1990partial, stone1990continuum}):
\begin{align} \label{eq_pls_coe}
    \widehat{\mathbf{b}} = \hat{R}(\hat{R}^TS_{xx}\hat{R})^{-1}\hat{R}^Ts_{xy},
\end{align} 
where $\hat{R} = (s_{xy}, S_{xx}s_{xy}, \dots, S_{xx}^{q-1}s_{xy})$ is the $p \times q$ matrix of the Krylov sequence with a $p \times p$ matrix $S_{xx}$  and a $p \times 1$ vector $s_{xy}$ defined as follows:
\begin{align*}
    S_{xx} = \frac{\mathbf{X}^T(I - 11^T/N)\mathbf{X}}{N-1}, \\
    s_{xy} =  \frac{(\mathbf{X} - \mathbb{E}(\mathbf{X}))^T( \mathbf{y} - \mathbb{E}(\mathbf{y}))}{N-1},
\end{align*}
where $I$ is an identity matrix and $1$ is a matrix of ones. Intuitively, the algorithm searches for factors  that capture the highest variability in $\mathbf{X}$, and at the same time maximizes the covariance between $\mathbf{X}$ and $\mathbf{y}$. If the number of components equals the dimension of the covariates, $q = p$, the method is equivalent to the ordinary least squares (\citealp{helland1990partial}). 

\cite{phatak1997geometry} show that the partial least squares estimator can geometrically be interpreted as the tangent rotation and projection of the OLS estimator on the ellipsoid. Consequently, that property allows us to extract the dimensions of the feature space that are relevant for predicting the outcome as well as the policy effect heterogeneity.  
We determine the optimal number of target components based on the cross-validation results. Particularly, we choose the minimum number of components  beyond which the prediction performance stabilizes. It is noteworthy to emphasize that the target components identified through partial least squares are inherently derived from the data. This ensures that the composition of the identified groups may differ depending on the specific dataset under investigation. However, an article by \cite{cao2018partial} provides compelling evidence demonstrating the robustness of the partial least squares method even in the presence of potential misspecifications in the treatment assignment model.

\subsection{Estimation of Personalized Policy Effects} \label{eq_causal_imp}

To identify smaller subgroups within the population, we use a causal forest algorithm. A tree in causal forests recursively partitions the feature space, in this setting, the space of identified target components $\mathbf{C}$, and makes axis-aligned splits to estimate the conditional mean of the outcome $\mu(c) = \mathbb{E}(y_{i}|P_i, \mathbf{C}_i = c)$ at a point $c$ for $P_i \in \{0, 1\}$.

An axis-aligned split is a pair $m = (j, c)$, where $j = 1, \dots, q$ is a specific component (the \textit{splitting coordinate}) and $c \in \mathbb{R}$ is the corresponding value (the \textit{splitting index}). The recursive partitioning procedure begins by considering the set $\mathcal{P}^{(0)} = \mathbf{C}\in R^{q}$ (the \textit{parent node} of the tree). For this set, we select the splitting coordinate $j: 1 \leq j \leq q$ and the splitting index $c$ that divide $\mathcal{P}^{(0)}$ into two non-overlapping rectangles (\textit{child nodes}): 
\begin{align}
    \mathcal{P}^{(1, 1)} = \mathcal{P}^{(0)} \cap \{\widetilde{c}\in \mathcal{P}^{(0)}: \widetilde{c}_j \leq c\} \ \text{and} \ \mathcal{P}^{(1, 2)} = \mathcal{P}^{(0)} \cap \{\widetilde{c}\in \mathcal{P}^{(0)}: \widetilde{c}_j > c\},
\end{align}
After the first split, the process is repeated for $\mathcal{P}^{(1, 1)}$ and $ \mathcal{P}^{(1, 2)}$separately until the desired level of partitioning is achieved.

The sequence of $k$ splits defines a partition of the component space $\mathbf{C}$, which we denote by $\Pi$. This partition (or equivalently, a tree) consists of non-overlapping rectangular regions $\ell_n$ called the \textit{leaves} or \textit{terminal nodes} of the tree. These leaves represent the final subgroups or subpopulations identified by the algorithm. The union of all these partitions is the entire component space:
\begin{align*} 
\label{eq_tree} \Pi = \{\ell_1, \ell_2, \dots,  \ell_{|\Pi|}\} \ \  \text{and}  \  \  \cup_{n=1}^{|\Pi|}\ell_n = \mathbf{C}.
\end{align*}

\cite{athey2016recursive} propose a method for estimating heterogeneous policy effects under the assumption of unconfoundedness. To implement this method, we split the data into two different samples: a training sample $S^{tr}$ used to build and find the splitting variables and values, and an estimation sample $S^{est}$ used to estimate policy effects across different subgroups of the population. The unbiased sample analogue of $\mathbb{E}(\theta(\mathbf{C}_i))$ is denoted as follows:

\begin{align} 
    \widetilde{\theta}(\mathbf{C}_i, S^{est}, \Pi)  = \nonumber \sum_{n=1}^{|\Pi|} & \bigg( 1(c \in \ell_n, P_i = 1)\frac{1}{|i: \mathbf{C}_i \in \ell_n, P_i = 1|}\sum_{i: \mathbf{C}_i \in \ell_n}Y_i(1) - \\
    &1(c \in \ell_n, P_i = 0)\frac{1}{|i: \mathbf{C}_i \in \ell_n, P_i = 0|}\sum_{i: \mathbf{C}_i \in \ell_n}Y_i(0)\bigg),
\end{align}  
where $|\Pi|$ is the total number of the terminal nodes. $1(c \in \ell_n, P_i = d)$ is a binary variable and equals one when, for a given $d \in \{0, 1\}$,  a generic test data point $c$ belongs to a terminal leaf $\ell_n$, and zero otherwise.  Additionally, let $\Sigma$ be the variance of $\widetilde{\theta}(X_i, S^{est}, \Pi)$. \footnote{While our analysis is based on a single policy variable and a single outcome, the proposed framework can handle multiple policy variables and outcomes with correlated coefficients.  } 

To estimate policy effects from the available data, we aim to maximize the variance of the policy effect estimator:

\begin{align} \label{eq_loss_theta_new}
    \hat{\theta}(\mathbf{C}_i, S^{est}, \Pi) = \arg\max_{\widetilde{\theta}} \frac{1}{N^{tr}} \sum_{\ell} N^{tr}_{\ell}\widetilde{\theta}(\mathbf{C}_i, \Pi)^{T}\widehat{\Sigma}^{-1}\widetilde{\theta}(\mathbf{C}_i, \Pi).
\end{align}

The proof of \eqref{eq_loss_theta_new} is provided in Appendix \ref{app_loss}. Intuitively, the objective function in \eqref{eq_loss_theta_new} encourages the causal forest algorithm to search for subsets of target components with the highest variation in policy effects. To further increase the robustness of the estimates, we build multiple trees on bootstrapped data and average the resulting coefficients. This approach, known as the causal forest algorithm, has been described in detail by \cite{athey2016recursive} and \cite{wager2018estimation}.

\section{Assumptions and Large Sample Properties}
To show the asymptotic normality of the estimated policy effects, we need to make certain assumptions about the underlying data-generating process.

\begin{assump}[Data Generating Process]\label{ass_dgp_1}
    Let $\mathbf{y} = g(b_0 + \mathbf{X}\mathbf{b}) + \varepsilon$ where $\mathbf{b}$ is a $p \times 1 $ vector of coefficients, $b_0$ is a constant and $g$ is a non-linear mapping. Assume, $\mathbf{X}$ have a joint Elliptical distribution with  the mean $\mu_{\mathbf{X}}$ and a variance $\Sigma_{\mathbf{X}\mathbf{X}}$. Assume $\mathbf{X}$ is independent of $\varepsilon$.  Moreover, let $S_{xx}$ and $s_{xy}$ converge in probability to $\Sigma_{\mathbf{X}\mathbf{X}}$ (the population variance of $\mathbf{X}$) and $\sigma_{\mathbf{X}\mathbf{y}}$ (the population covariance of $\mathbf{X}$ and $\mathbf{y}$) when $N \rightarrow \infty$. Moreover, let there exist a pair of eigenvectors and eigenvalues $(v_j, \lambda_j)$ for which $\sigma_{\mathbf{X}\mathbf{y}} = \sum_{j=1}^M\gamma_jv_j$ (with $\gamma_j$ non-zero for each $j = 1, \dots, M$).  Assume also $\mathbb{E}(|g(U)|) < \infty$ and $\mathbb{E}(U|g(U)|) < \infty$ with $U = b_0 +  \mathbf{X}\mathbf{b}$ and q = M. 
\end{assump}

Under Assumption \ref{ass_dgp_1}, the relation between the response and the independent characteristics follows a predetermined functional form. Additionally, the subject characteristics are assumed to have an elliptical distribution, meaning they are shaped like an ellipse in a multi-dimensional coordinate system. While this assumption is not always satisfied in practice, it has been shown that the results obtained under this assumption do not significantly differ from those obtained when the features have other types of distributions (see \citealp{brillinger2012generalized}).

\begin{lemma} \label{lemma_pls_consistency}
Let Assumption \ref{ass_dgp} hold. Then $\hat{\mathbf{b}}$ in \eqref{eq_pls_coe}  is consistent up to a proportionality constant. 
\end{lemma} 
The proof of Lemma \ref{lemma_pls_consistency} is provided in Appendix \ref{app_brillinger}. Lemma \ref{lemma_pls_consistency} shows that the identified target components are consistent. The causal forest method described in this article relies on the same assumptions as those introduced by \cite{wager2018estimation}. One of them is the "honesty" of the tree.

\begin{assump}[Honesty] \label{ass_honesty}
The outcome $y_i$ and the splitting parameters (the splitting coordinates and indices, $m = (j, c)$) are independent of each other, conditional on the observed components $\mathbf{C}_i$. This independence holds for each subject $i$ whose outcome $y_i$ is used in the final prediction:
\begin{align*}
    F(y_{i}|\mathbf{C}_i, m) = F(y_{i}|\mathbf{C}_i). 
\end{align*}
$F$ denotes the density of the outcome variable.\footnote{If we have access to multiple outcomes, this assumption holds for each one individually. } 
\end{assump} 

There are various ways to satisfy Assumption \ref{ass_honesty}. In this article, we use a two-sample approach, where we split the data into a training sample $S^{tr}$ and an estimation sample $S^{est}$. The splitting coordinates and indices ($m$) of the trees are determined based on the observations in $S^{tr}$, while the predicted outcomes are based on the observations in $S^{est}$. This separation of the data into two different samples ensures that the splitting parameters and the outcomes are independent of each other.

\begin{assump}[Random Split Trees]\label{ass_randsplit}
At each recursive step, the probability of choosing the $j$-th component as the splitting coordinate is lower bounded by $\pi/d$ for $\pi \in (0, 1]$ and for all $j = 1, \dots, q$.
\end{assump}

In order to guarantee the consistency of the causal forest method, it is necessary for the leaves of the trees to become small in all dimensions of the component space as the sample size $N$ increases. To ensure this, we adopt Assumption \ref{ass_randsplit}, which is based on the assumptions of \cite{meinshausen2006quantile} and \cite{wager2018estimation}. This assumption states that for all splitting steps, each component has a probability of at least $\pi/d$ of being selected as the splitting coordinate, for some $0 < \pi \leq 1$. 

\begin{assump}[The Splitting Algorithm is ($\alpha, k$)-regular]\label{ass_regul}
There exists a positive constant $\alpha$ such that at each split, at least a fraction $\alpha$ of the available training examples are left on each side of the split. Additionally, we require that the splitting process ceases at a node when it contains less than $k$ observations for some $k\in \mathbb{N}$.
\end{assump}

Assumption \ref{ass_regul} ensures that each half-space produced by a split in the tree construction process contains a sufficient number of observations. As shown by \cite{wager2015adaptive}, this assumption also implies that the half-spaces are large in Euclidean volume. Assumption \ref{ass_regul} places an upper bound on the number of observations that can be contained in a terminal node of the tree. Specifically, when a tree is fully grown to depth $k$, we have that each terminal node contains between $[k, 2k-1]$ observations. One important consequence of this assumption is that it places an upper bound on the variance of the tree estimator at any test point $c$.

\begin{assump}[Distributional Assumptions on the Data Generating Process]\label{ass_dgp}
The target components $\mathbf{C}_i$ are supported on the unit cube $\mathbf{C}_i\in [0, 1]^p$, and the density of these components is bounded away from zero and infinity. The first and second moments of the outcome, $\mathbb{E}(y_{i}|\mathbf{C}_i = c)$ and $\mathbb{E}\big(y_{i}^2|\mathbf{C}_i = c\big)$, are Lipschitz-continuous functions of the target components.
The variance of the outcome, $Var(y_{i}|\mathbf{C}_i = c)$, is bounded away from zero for all values of the target components. Specifically, we have $inf_{c \in \mathbf{C}} Var(y_{i}|\mathbf{C}_i = c) > 0$. 
\end{assump}

Lipschitz continuity and bounded variances are widely used assumptions in the field of statistics and machine learning (\citealp{wager2018estimation, biau2012analysis}). In the context of this paper, the results do not depend explicitly on the distributional assumptions of $\mathbf{C}_i$, however, they affect the constants that we carry throughout this paper (constants borrowed from Lemma 2 and Theorem 3 in Section 3.2 in \citealp{wager2018estimation}).  

\begin{assump}[\textbf{Overlap}]\label{ass_overlap}
Let $0 < \epsilon < 1$, and consider any element $c \in [0, 1]^q$. Then the following holds:
$$
\epsilon < \mathbb{P}(P_i = 1 \mid \mathbf{C}_i = c) < 1 - \epsilon.
$$
\end{assump}

Assumption \ref{ass_overlap} ensures that, as the number of observations $N$ increases, there will be a sufficient number of subjects with and without a policy at any given test point $c$. 
Under the given assumptions, \cite{wager2018estimation} show that the random forest estimator is consistent and asymptotically normally distributed. They generalize the properties to a single parameter of interest. Additionally, \cite{nareklishvili2022adaptive} in Theorem 6.3 shows that the causal forest estimator is asymptotically normally distributed even for multiple (possibly correlated) parameters. Appendix \ref{app_main_asymp} presents supplementary definitions and elaborates on the algorithm. To quantify the uncertainty of the policy effects, we employ the jackknife variance estimator, as outlined in Subsection \ref{app_sub_inference} of Appendix \ref{app_main_asymp} (\citealp{wager2018estimation}). The  results of \cite{wager2018estimation} and \cite{nareklishvili2022adaptive} directly apply to our setting when conditioned on the target components.

\section{Simulated Experiments}
\begin{figure}[H]
	\centering
	\includegraphics[width = \linewidth]{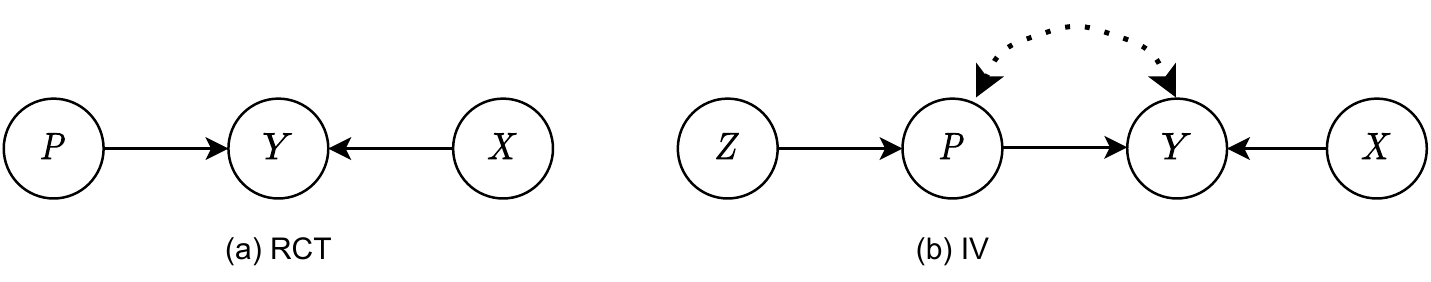}
	\caption{(a) Randomized controlled trial (RCT). In this design, independent variable(s) X affect the outcome $Y$, but not the policy $P$. (b) Instrumental variable design (IV). Within this framework, an instrumental variable (an exogenous variable denoted by $Z$) affects the policy $P$ but does not have a direct influence on the outcome $Y$. In this setting, unobserved confounders (represented by the dashed arrows) are present and correlate with the policy and outcome. }\label{fig_chains}
\end{figure}

We explore different simulation designs aimed at evaluating the estimation performance of Forest-PLS. The accompanying Figure \ref{fig_chains} visually represents two distinct scenarios. In Panel (a), we present a randomized controlled experiment. In this scenario, the outcome of interest is influenced by a specific policy, and the assignment of the policy is under the complete control of the experimenter. This design allows us to assess the direct impact of the policy on the outcome while minimizing potential confounding factors. Panel (b) highlights an alternative situation where unobservable factors, which are not accessible to a policy-maker, may correlate with both the policy and the outcome.  This design aims to mimic real-world scenarios where policy decisions are made under uncertainty, and there exist latent factors that introduce bias in the analysis.

\subsection{Randomized Controlled Trials}
In the framework of randomized controlled trials (RCT), the policymaker has access to the outcome, denoted as $Y_i$, the policy $P_i$, and four distinct variables, represented as $X_{i1}$, $X_{i2}$, $X_{i3}$, and $X_{i4}$. Formally, the outcome ($Y_i$) is determined by a combination of the policy intervention and the individual characteristics associated with each participant $i = 1, \dots, N$:  
\begin{align} 
    \text{(a) RCT} & \nonumber \\
    &X_{i1}, X_{i2}, X_{i3}, X_{i4}, \sim \mathcal{N}(N, \mu, \Sigma), \ \mu = (-1, 1, 2, 0), \ \Sigma = \mathbf{1}_{4 \times 4}, \nonumber \\
    &P_i \sim  \mathcal{B}(N, 1, 0.5), \  \varepsilon_i \sim \mathcal{N}(N, 0, 1) \nonumber \\
    & Y_i  =  100\times X_{i1} + 100\times X_{i2} + P_i\times (X_{i3} + 0.1\times X_{i4} + 0.2\times X_{i3}\times X_{i4}) + \varepsilon_i.\label{eq_rct_v2}
\end{align}
In our paper, the policy variable ($P_i$) is generated from a binomial distribution, denoted as $\mathcal{B}$. This variable represents the policy intervention assigned to each participant and captures the binary nature of the treatment.
On the other hand, the individual characteristics ($X_{i1}$, $X_{i2}$, $X_{i3}$, and $X_{i4}$) are drawn from a normal distribution, denoted as $\mathcal{N}$. These characteristics could correspond to important demographic factors such as age, gender, education level, and race, and are continuous in nature. The outcome also consists of a normally distributed noise ($\varepsilon_i$). To investigate the effects of the policy intervention on the outcome, we simulate the randomized controlled trial (RCT) fifty times with different sample sizes denoted by $N$ = (70, 100, 500, 1000, and 5000), and average the results. The variance-covariance matrix of the participant characteristics ($\Sigma$) is assumed to be the identity matrix $\mathbf{1}_{4 \times 4}$, indicating that the individual characteristics are independent and have equal variances.


The randomized controlled trial design in \eqref{eq_rct_v2} presents two significant challenges for the Forest-PLS algorithm. First, the policy effects exhibit heterogeneity and vary across individuals based on their values of $X_{i3}$ and $X_{i4}$, i.e., $\mathbb{E}(Y_i|P_i = 1) - \mathbb{E}(Y_i|P_i = 0)  = X_{i3} + 0.1\times X_{i4} + 0.2\times X_{i3}\times X_{i4}$.  Second, the outcome is heavily influenced by independent variables $X_{i1}$ and $X_{i2}$, which are not particularly relevant for explaining the heterogeneity of policy effects. In this setting, the causal forest needs to correctly identify the sources of heterogeneity when provided with the target components.

\begin{figure}[H]
	\centering
	\subfloat[N = 70]{\includegraphics[width=0.5\linewidth]{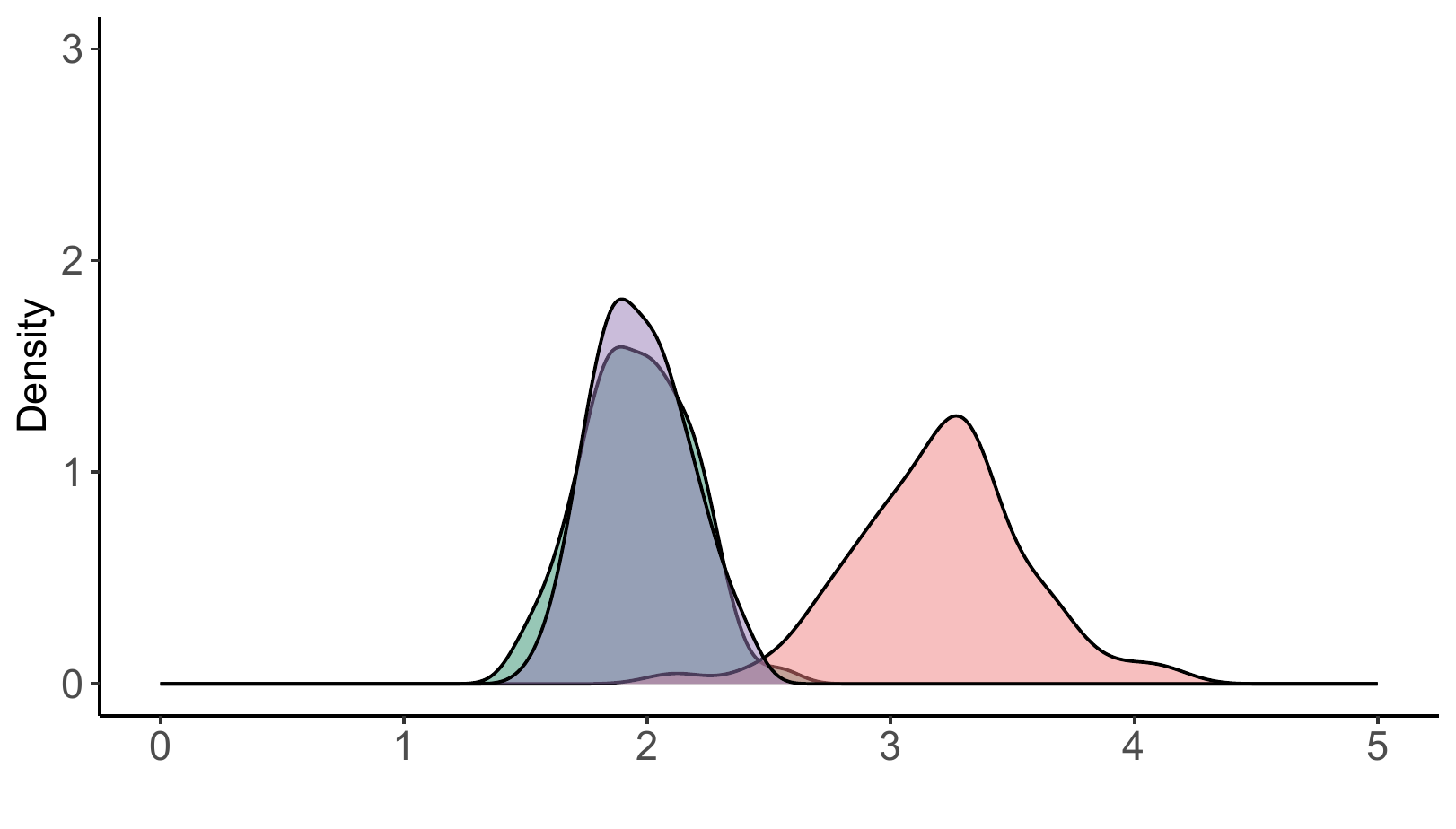}} 
   \subfloat[N = 500] {\includegraphics[width=0.5\linewidth]{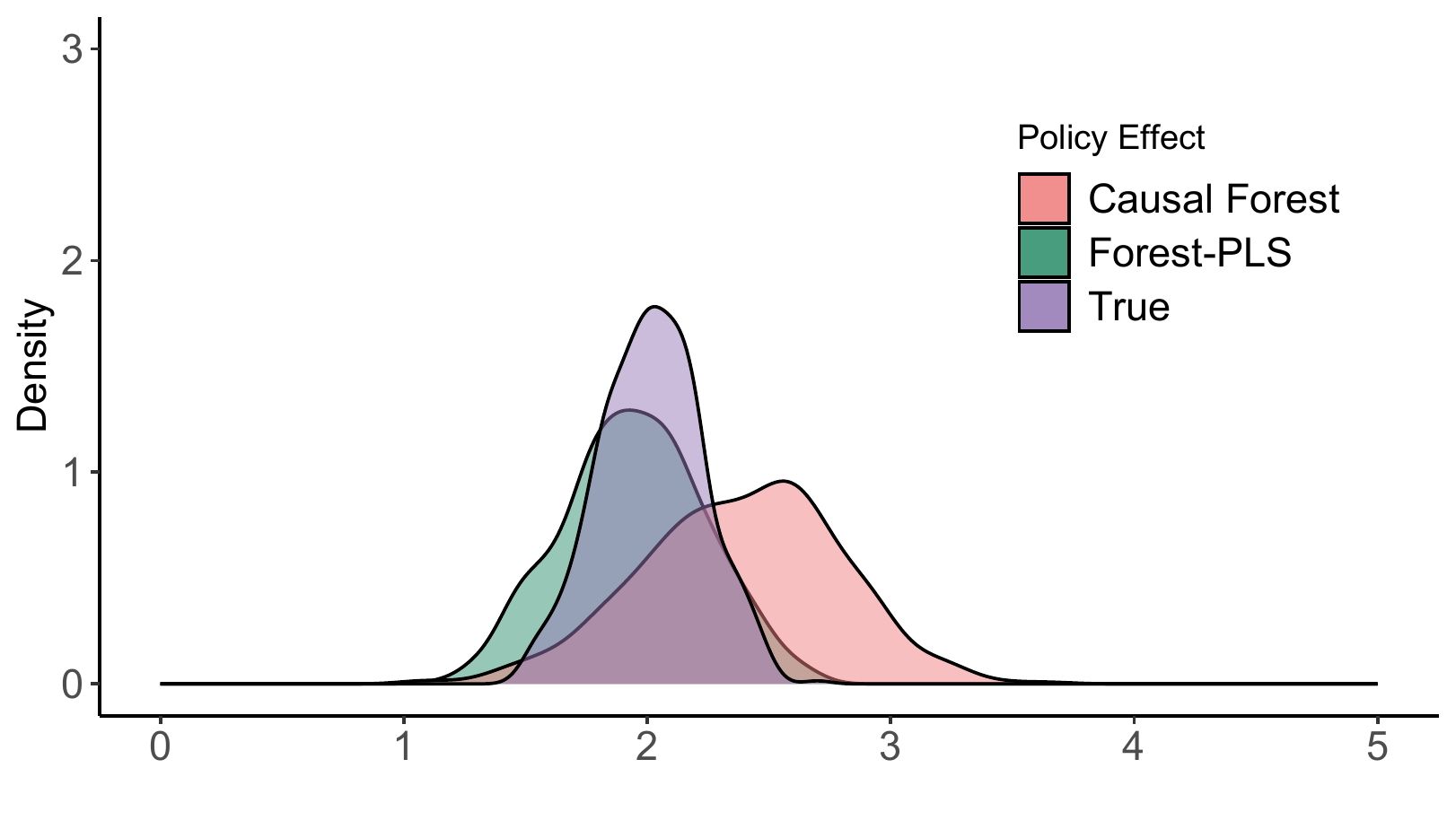}} \quad
	\subfloat[N = 1000]{\includegraphics[width=0.5\linewidth]{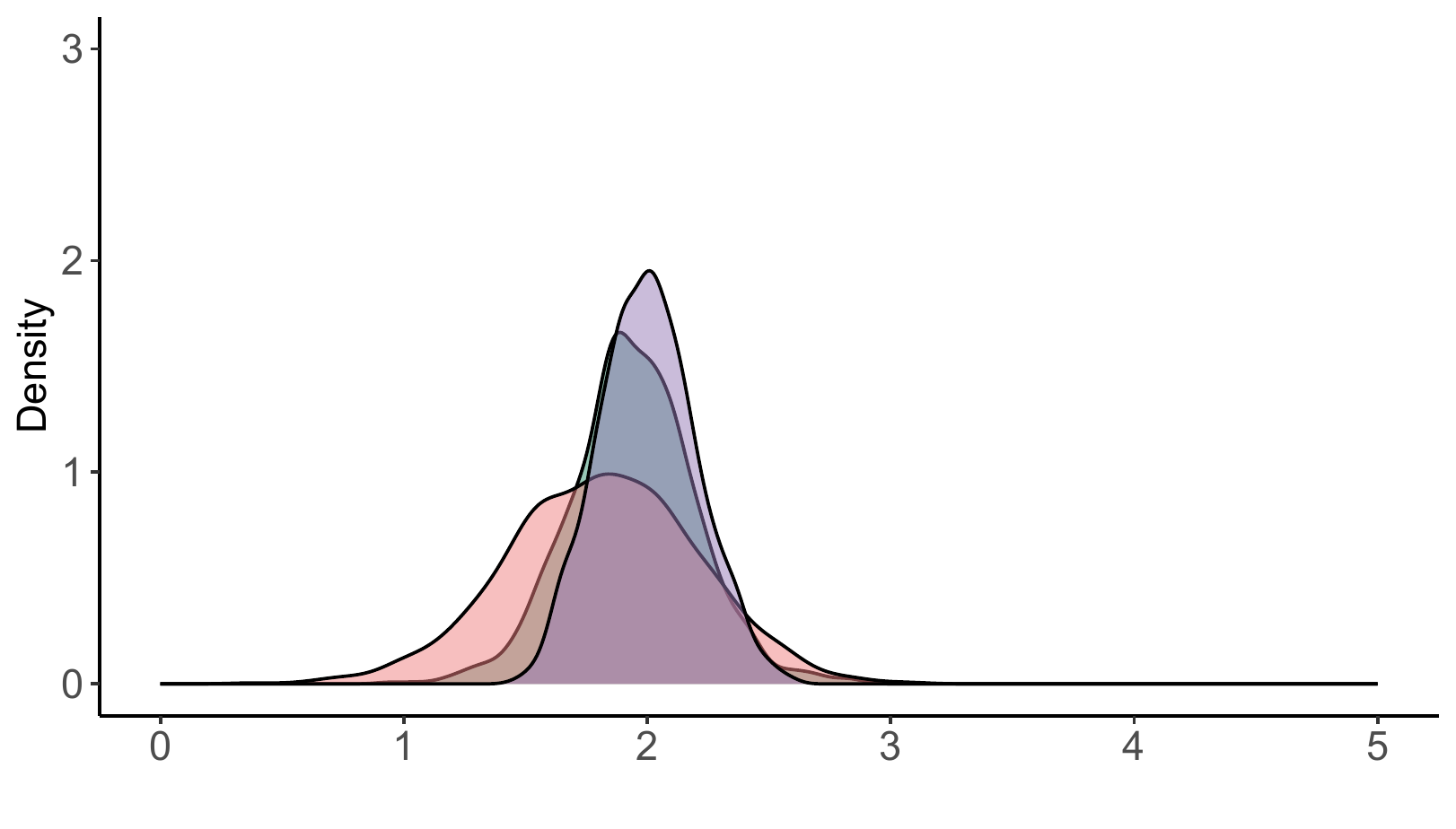}}
    \subfloat[N = 5000] {\includegraphics[width=0.5\linewidth]{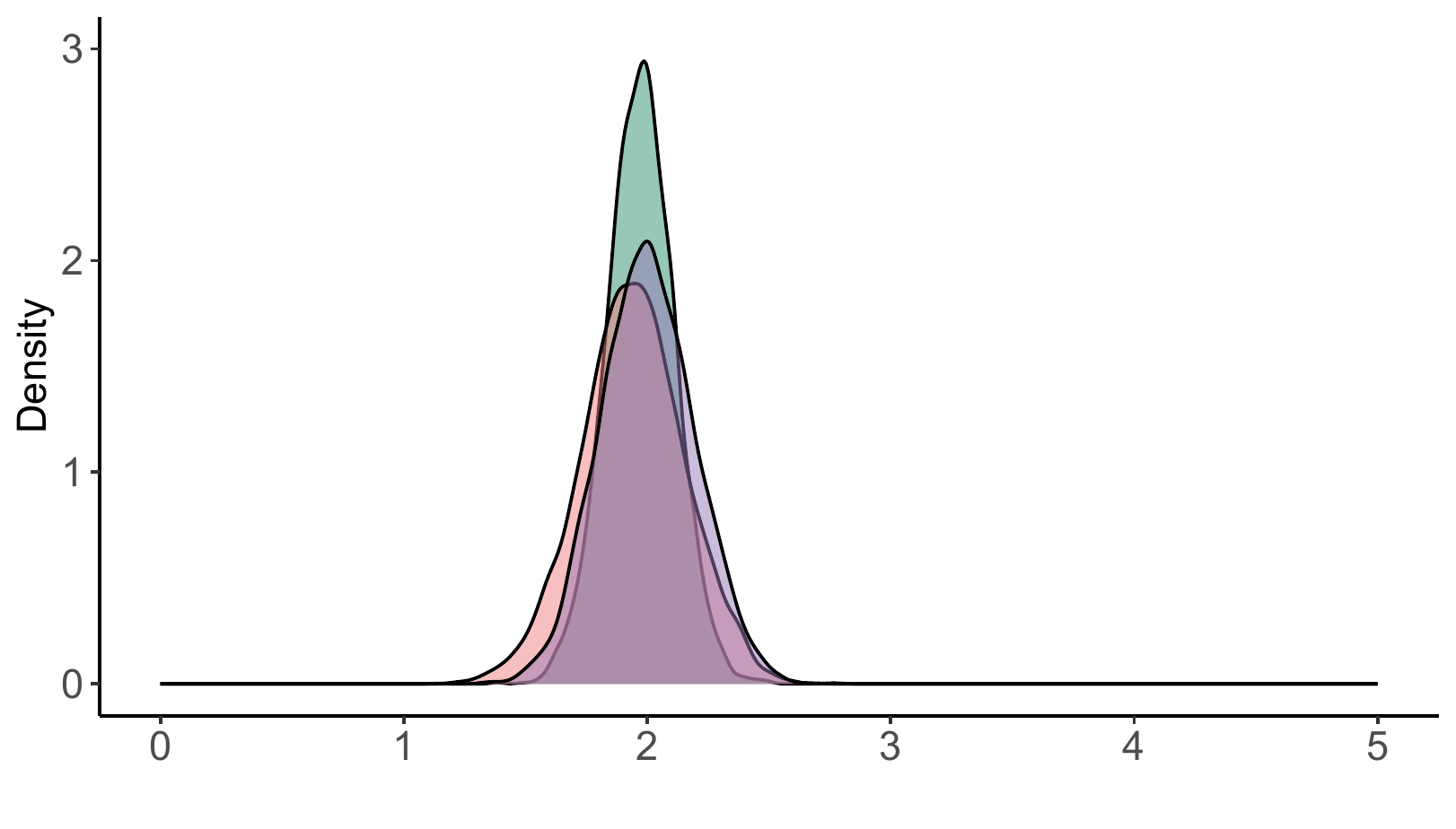}} \quad 
	\caption{Simulated experimental analysis: impact of varying observation counts on the performance of Forest-PLS. Each experiment comprises 1000 trees. Forest-PLS incorporates two optimal target components selected based on five-fold cross-validation results. The presented distributions represent the averages derived from fifty distinct replications of the experiment.  }\label{fig_group_heterogeneity}
\end{figure}

Figure \ref{fig_group_heterogeneity} depicts the simulated density of policy effects alongside their estimated counterparts based on the causal forest and Forest-PLS algorithms. The illustration reveals two notable advantages of Forest-PLS over the causal forest algorithm. First, Forest-PLS exhibits remarkable resilience in recovering the true density of policy effects, even when confronted with a limited number of observations. This characteristic renders the algorithm highly robust and consistent across varying sample sizes. Second, Forest-PLS effectively mitigates the influence of redundant variables and noise present in the experimental setup, enabling accurate estimation of the variance of policy effects.

The success of Forest-PLS can be attributed to the inherent diversity and relevance of the target components.  As demonstrated in Table \ref{tab_tc} in Appendix \ref{app_contr}, the first component predominantly captures the features $X_{i1}$ and $X_{i2}$, while the second component encapsulates the remaining information, $X_{i3}$ and $X_{i4}$. Each target component contains information valuable for explaining either the outcome or the policy effect heterogeneity. Thus, when trained on the target components, the causal forest efficiently retrieves the correct dimensions of the policy effect heterogeneity even in small samples. Table \ref{tab_varimp} in Appendix \ref{app_contr} verifies that the variable importance for policy effect heterogeneity is similar for both methods. We further compare partial least squares with LASSO regression. Table \ref{tab_lasso} in Appendix \ref{app_contr} shows that, unlike partial least squares, LASSO drops the variables that are essential for explaining policy effect heterogeneity. 
Figures \ref{fig_group_heterogeneity_noconf} and \ref{fig_group_heterogeneity_nbd} in Appendix \ref{app_additional_sims} provide additional simulation designs and show the advantages of Forest-PLS over the conventional causal forest method.

\subsection{Observational Data}
In the second simulated experiment, we aim to assess the variation in policy effects in the presence of an unobservable latent variable,  denoted as $U_i$. This latent factor impacts both the policy implementation and the outcome. Furthermore, the policy itself is influenced by two instrumental variables, namely $Z_{i1}$ and $Z_{i2}$: 
\begin{align} \label{eq_iv}
    \text{(b) IV} &\nonumber  \\
    &X_{i1}, X_{i2}, X_{i3}, X_{i4}, \sim \mathcal{N}(N, \mu, \Sigma), \ \mu = (-1, 1, 2, 0),  \ \Sigma = \mathbf{1},  \nonumber \\
    & U_i \sim \mathcal{N}(N, 0, 1),  \nonumber  \\
    & \varepsilon_{i1} \sim \mathcal{N}(N, 0, 1), \ \varepsilon_{i2} \sim \mathcal{N}(N, 0, 1)\nonumber  \\
    & P_i = Z_{i1}- Z_{i2} + 0.5\times U_i + \varepsilon_{i1}, \nonumber \\ 
    & Y_i =  0.5\times P_i\times X_{i1} - 3\times U_i + \varepsilon_{i2}.
\end{align}

The policy and outcome consist of errors $\varepsilon_{i1}$ and $\varepsilon_{i2}$, respectively. In \eqref{eq_iv}, the policy effect is represented by $0.5\times X_{i1}$. The framework in \eqref{eq_iv} highlights the presence of a weak instrument and a strong confounder ($U_i$). Unlike the RCT design, in this framework, we have a single variable relevant to explaining the outcome and policy effect heterogeneity. We assume, the researcher does not have access to either the instruments $Z_{i1}$, $Z_{i2}$, or the latent variable $U_i$. Therefore, the algorithm receives the observable characteristics  $X_{i1}, X_{i2}, X_{i3}, X_{i4}$ as the input. 

\begin{figure}[H]
	\centering
	\subfloat[N = 70]{\includegraphics[width=0.5\linewidth]{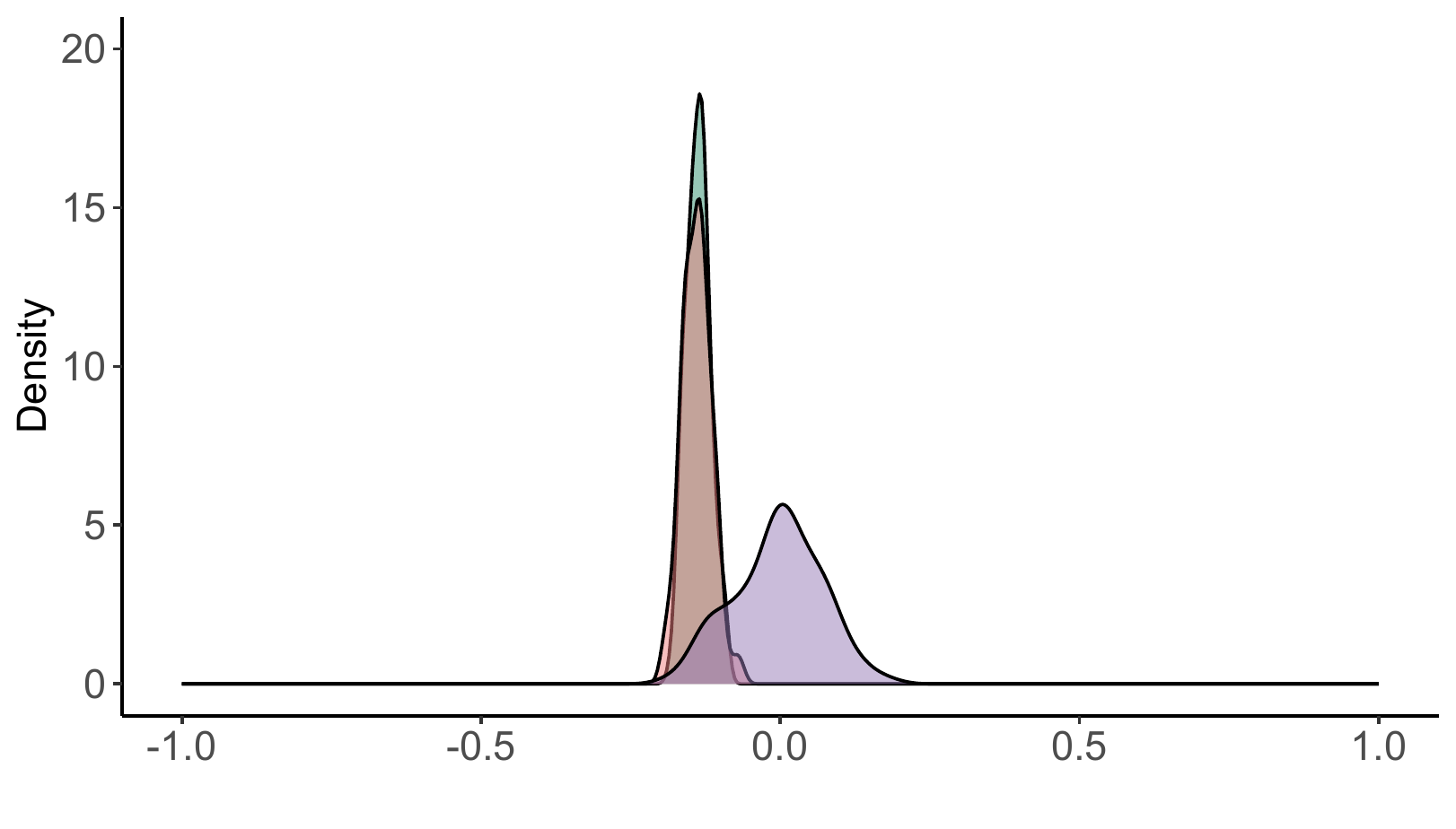}} 
   \subfloat[N = 500] {\includegraphics[width=0.5\linewidth]{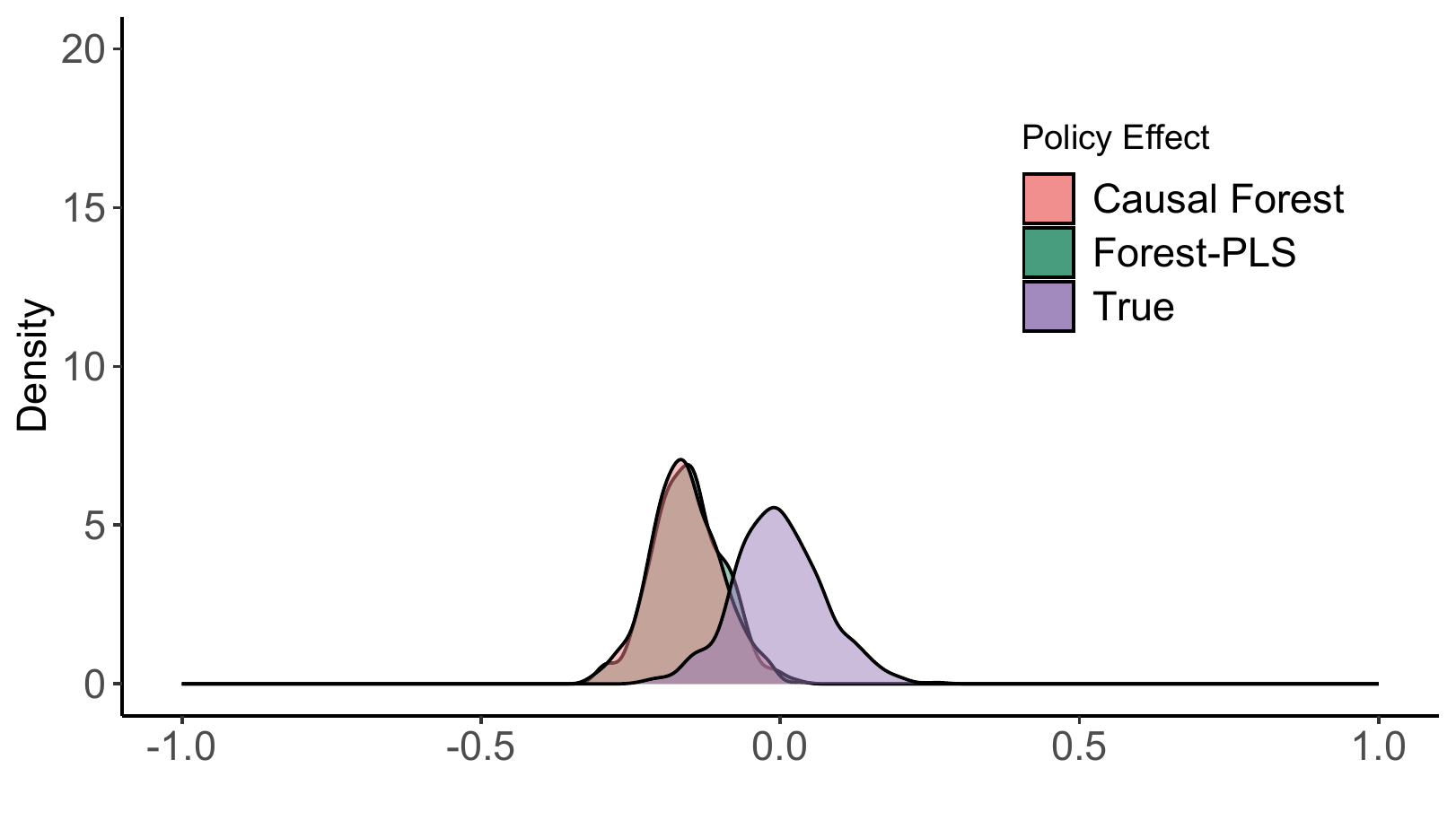}} \quad
	\subfloat[N = 1000]{\includegraphics[width=0.5\linewidth]{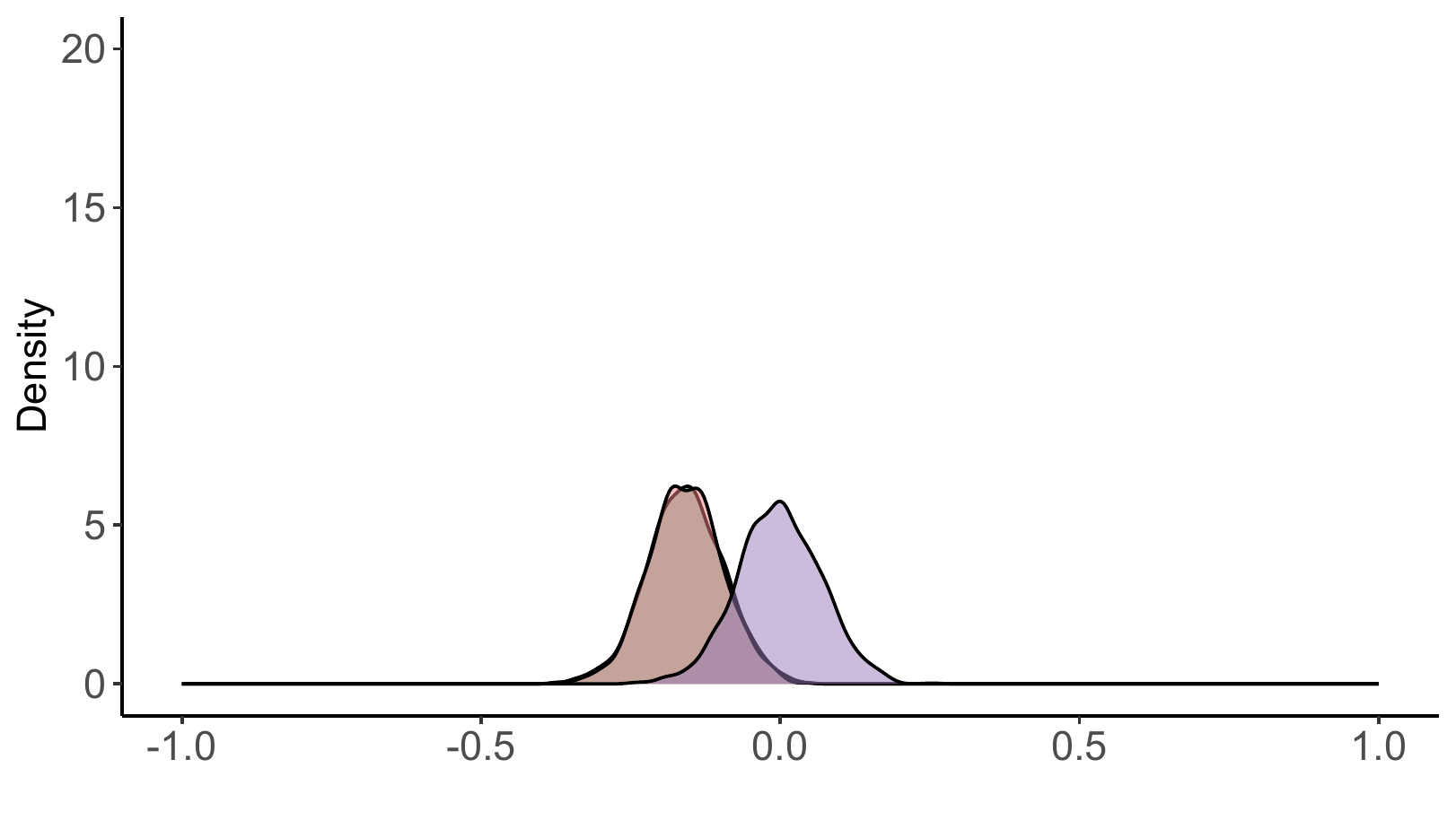}}
    \subfloat[N = 5000] {\includegraphics[width=0.5\linewidth]{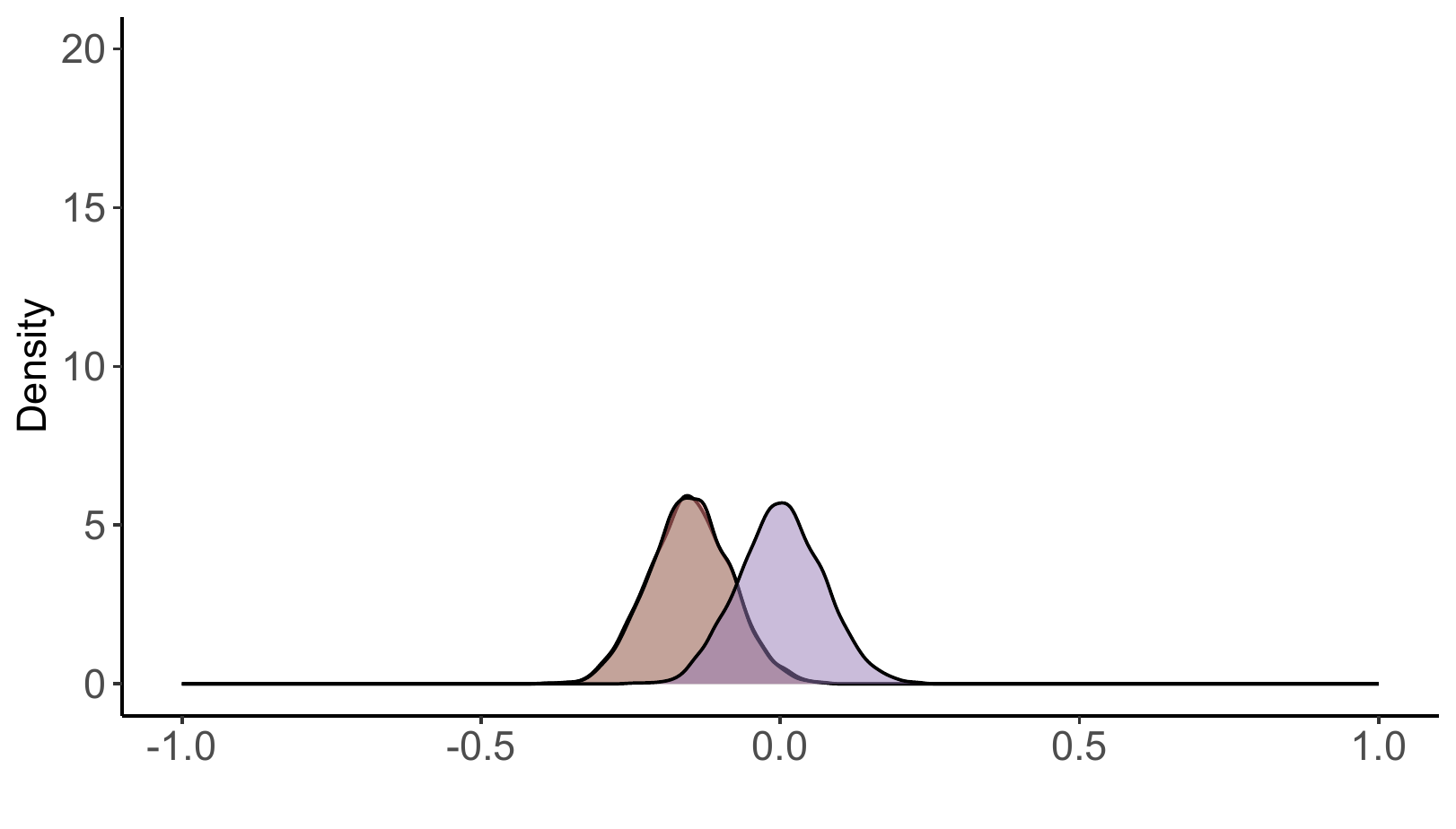}} \quad 
	\caption{Simulated experiments for a varying number of sample sizes. In each experiment, the number of trees equals 1000. Forest-PLS consists of two optimal target components chosen based on five-fold cross-validation results (the optimal number of components is the minimum number beyond which the cross-validation root-mean-squared-error stabilizes). Each method receives original features as the input, denoted as $X_{i1}$, $X_{i2}$, $X_{i3}$, and $X_{i3}$, wherein three of these variables are redundant. The presented distributions represent the averages derived from fifty distinct replications of the experiment.  } \label{fig_liv}
\end{figure}

Figure \ref{fig_liv} illustrates that the Forest-PLS and causal forest algorithms systematically and identically underestimate the mean and variance of the policy effects. We also find that the methods are comparable across a varying number of observations. 

\section{Reemployment Experiment in Pennsylvania}
The Pennsylvania "Reemployment Bonus" Demonstration was a randomized controlled trial in 1988-89 that aimed to investigate the impact of financial incentives on the reemployment outcomes of unemployed individuals. The study population was divided into a control group, which received the usual benefits provided by the Unemployment Insurance System, and six treatment groups. Treated individuals were offered a cash bonus for fulfilling certain criteria related to finding and retaining employment. Specifically, participants in the treatment groups were required to accept a bonus that would be paid to them if they were able to secure a full-time job of at least 32 hours per week within a specified period (the qualification period) and maintain that employment for at least 16 weeks.

Two bonus levels were tested. These two levels were a low bonus and a high bonus, which were respectively three and six times the weekly benefit amount (WBA) received by the participants. The low bonus was on average $\$500$, while the high bonus was $\$997$. In addition to these two levels of bonus, the study also considered two different qualification periods, starting from the date on which the bonus offer was made. These periods were a short one of 6 weeks and a long one of 12 weeks.

In addition to testing the impact of financial incentives on reemployment outcomes, the Pennsylvania "Reemployment Bonus" Demonstration also aimed to investigate the effectiveness of providing job-search assistance to unemployed individuals. To this end, participants in the treatment groups were offered a workshop and an individualized assessment session as part of the treatment design. However, attendance at the workshop and completion of the assessment session were not mandatory for claimants.

In this article, we focus on treatment Group 4 which received a high bonus amount and a long qualification period, as well as an offer of a workshop. The primary outcome of interest is the logarithm of unemployment duration in weeks. The data include twenty different characteristics of the claimants, such as age, gender, the quarter of the experiment in which they enrolled, and unemployment rates in the local area. \footnote{The variables are described in detail at the following url: \url{http://qed.econ.queensu.ca/jae/2000-v15.6/bilias/readme.b.txt}.} Further information about the experiment and data can be found in the article by \cite{bilias2000sequential}.


\subsection{Target Components}
In this section, we identify and characterize policy-relevant target components. According to Figure \ref{fig_cv} in Appendix \ref{app_expl}, the optimal number of components equals two. To characterize and interpret the chosen scores, Table \ref{app_table_comp} in Appendix \ref{app_expl} illustrates the effect of the claimant characteristics on each target component. The negative value of a coefficient indicates that there is an inverse relationship between the characteristic in question and the outcome being measured. For instance, a black claimant is associated with a 1.3\% lower score on average relative to a white claimant. Based on the sign of the coefficients in Table \ref{app_table_comp}, we can interpret the scores.  

\begin{align}\label{eq_comp1}
    \text{Low score values} \sim
   \begin{Bmatrix} 
      \text{Young} \ (\text{age} <= 35) \\ 
      \text{Male} \\
      \text{Non-white}\\
        \text{Joined the experiment early}\\
      \text{In the sector of non-durable manufacturing}\\
      \text{Few or no dependents}\\
   \end{Bmatrix} 
\end{align}

\begin{align}\label{eq_comp2}
    \text{High score values} \sim
   \begin{Bmatrix} 
      \text{Middle-age and older}  \\ 
      \text{Female} \\
      \text{White}\\
        \text{Joined the experiment late}\\
      \text{In the sector of durable manufacturing}\\
      \text{Many dependents}
   \end{Bmatrix} 
   \end{align}

The lowest values of the target components identified in this article correspond to a subgroup of young, non-white male claimants in the non-durable manufacturing sector who enrolled in the experiment early on. On the other hand, the highest values of the components reflect a subgroup of middle-aged and older female individuals in the durable manufacturing sector. These claimants enrolled in the experiment late in the final quarter and tend to have a high number of dependents (as indicated by the positive coefficient for the "dep" variable in Table \ref{app_table_comp} of Appendix \ref{app_expl}).  \eqref{eq_comp1} and \eqref{eq_comp2} summarize the characteristics of subgroups. \footnote{According to Table \ref{app_table_comp} in Appendix \ref{app_expl}, these two components are almost identical in this setting, therefore, \eqref{eq_comp1} and \eqref{eq_comp2} hold for each. } Components are continuous, therefore, they characterize a full spectrum of individuals from one group to another.

\subsection{Effect Heterogeneity}
\begin{figure}[H]
\centering
\begin{tabular}{cc}
\includegraphics[width = \linewidth]{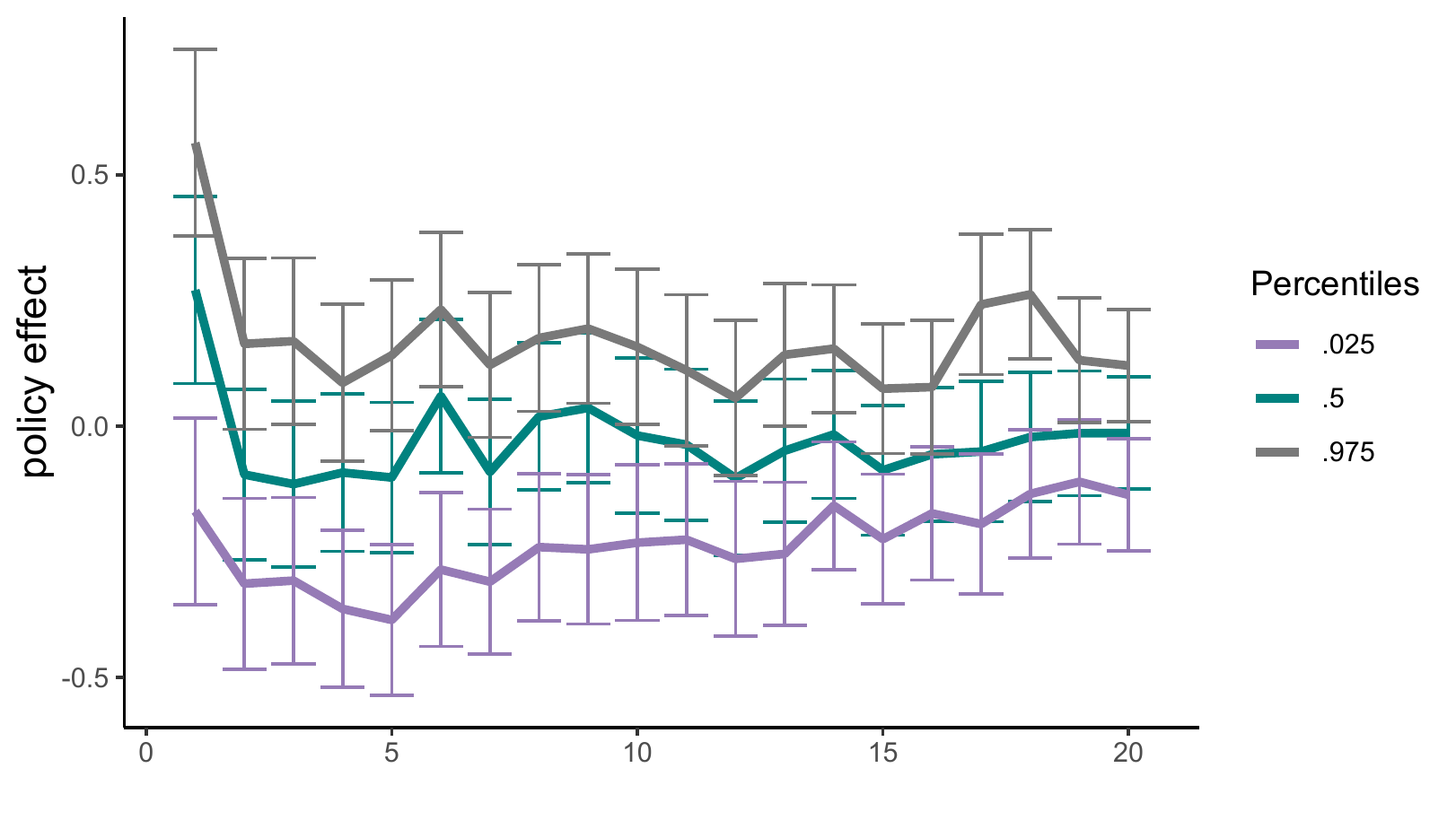} \\
(a) Component 1 \\
\includegraphics[width = \linewidth]{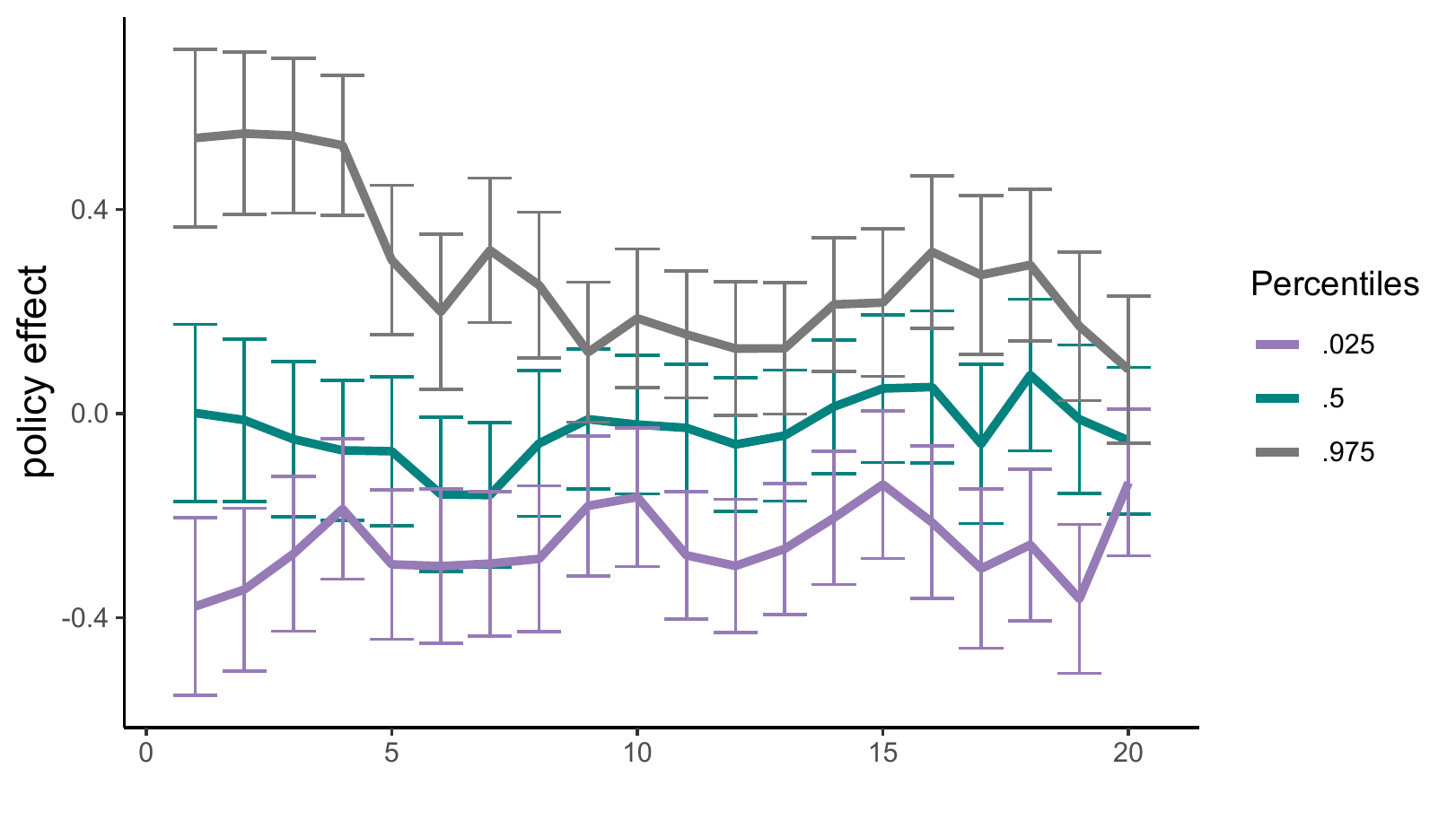} \\
(b) Component 2
\end{tabular}
\caption{x-axis represents vigintiles of the corresponding component, while y-axis measures the "Reemployment Bonus" effect on the logarithm of unemployment duration in weeks. The plot shows the error bars with the color corresponding to a given percentile of the policy effect. The number of trees equals 1000. See Appendix \ref{app_expl} for a detailed description of the optimal number of target components. }
\label{fig_het_c2}
\end{figure}

In this section, we investigate the heterogeneity, or variability, in the effect of financial incentives on unemployment duration within and across different values of the components. This is done by examining various percentiles of the reemployment bonus effect on unemployment duration across the corresponding component vigintiles, as shown in Figure \ref{fig_het_c2}. 

The results depicted in Figure \ref{fig_het_c2} reveal considerable variation in the policy effects both within and between groups. In particular, financial incentives have been found to potentially motivate some young non-white individuals (the group represented by low score values in \eqref{eq_comp1} to enter the labor market. However, these incentives may also provide a temporary financial cushion for others, potentially dissuading them from actively seeking employment. In contrast, this variation is less pronounced among older white claimants (upper vigintiles of each component, the group corresponding to high score values in \eqref{eq_comp2}. 

To contrast our method with the causal forest algorithm, Figure \ref{fig_group_heterogeneity_cf} in Appendix \ref{app_expl} assesses the heterogeneity of policy effects based on the latter. Unlike Forest-PLS, the causal forest does not depict significant heterogeneity either across the target components, or the original independent characteristics.

Our analysis shows that the "Reemployment Bonus" policy has, on average, a negative effect on unemployment duration. However, significant variation is captured by different percentiles of policy effects. The difference between the 97.5th and 2.5th percentiles of policy effects is 92.8\% in the first vigintile of Component 2, and this difference decreases to 22.1\% in the 20th vigintile of the same component. These findings suggest the need for targeted, personalized measures for younger non-white male claimants.

\section{Conclusion}
Policymakers frequently seek to understand the impact of interventions on specific subgroups or segments of the population, defined by certain characteristics or attributes known as covariates. In this article, we present a method for analyzing the density of policy effects within these target segments, which are defined as linear combinations of the explanatory variables. To achieve this, we combine two existing techniques, partial least squares and causal forests, that allow us to identify and analyze policy effects for the full range of the target segments. This approach enables policymakers to understand how policy effects vary within and across these segments, providing valuable insights for personalized policy analysis.

We show that the method is consistent and leads to asymptotically normally distributed policy effects. Additionally, our approach generalizes beyond a single policy effect to multiple (plausibly) correlated policy effects. Our analysis based on data from Pennsylvania "Reemployment Bonus" Demonstration reveals a significant variation in the effect of financial incentives on the logarithm of unemployment duration. The findings highlight the need for targeted measures for young non-white male participants. 

One potential extension of our method is to incorporate quantile regression forests \citep{meinshausen2006quantile}, conditional on the target components. In addition to using randomized control trials, we also plan to explore the application of our method to observational data with an endogenous policy \citep{vella1999estimating, baiocchi2014instrumental}.  By applying the method to observational data, we hope to gain further insights into the heterogeneity of policy effects and inform the design of  interventions for institutional settings.

\bibliographystyle{plainnat}
\bibliography{output}

\appendix
\newpage
\markboth{Appendices}{}
\addcontentsline{toc}{chapter}{Appendices}
\renewcommand{\thesection}{A.\arabic{section}}

\section{The Loss Function}\label{app_loss}
Minimize the difference between the personalized and group-level parameters, 
\begin{align} \label{eq_loss}
   \arg\min_{\widetilde{\theta}(\mathbf{C}_i, S^{est})}  \mathbb{E}_{S^{tr}, S^{est}} & \bigg[\big(\theta(\mathbf{C}_i) - \widetilde{\theta}(\mathbf{C}_i, S^{est}, \Pi)\big)^T\Sigma^{-1}\big(\theta(\mathbf{C}_i) - \widetilde{\theta}(\mathbf{C}_i, S^{est}, \Pi)\big) - \theta(\mathbf{C}_i)^T\Sigma^{-1}\theta(\mathbf{C}_i)\bigg]  = \\
    \mathbb{E}_{S^{tr}, S^{est}}& \bigg[\underbrace{\big(\theta(\mathbf{C}_i) - \theta(\mathbf{C}_i, \Pi)}_{A} + \underbrace{\theta(\mathbf{C}_i, \Pi) -  \theta(\mathbf{C}_i, S^{est}, \Pi)}_{B}\big)^T\Sigma^{-1}\big(\underbrace{\theta(\mathbf{C}_i) - \theta(\mathbf{C}_i, \Pi)}_{A} + \nonumber\\
    & \underbrace{\theta(\mathbf{C}_i, \Pi) - \theta(\mathbf{C}_i, S^{est}, \Pi)}_{B}\big)- \theta(\mathbf{C}_i)^T\Sigma^{-1}\theta(\mathbf{C}_i)\bigg] = \\ \nonumber
    & \mathbb{E}_{S^{tr}}\bigg ( \theta(\mathbf{C}_i)^T\Sigma^{-1}\theta(\mathbf{C}_i) - 2\theta(\mathbf{C}_i)^T\Sigma^{-1}\theta(\mathbf{C}_i, \Pi) \  + \nonumber \\
    &\theta(\mathbf{C}_i, \Pi)^T\Sigma^{-1}\theta(\mathbf{C}_i, \Pi)  
    - \theta(\mathbf{C}_i)^T\Sigma^{-1}\theta(\mathbf{C}_i)\bigg)  \ + \\
    & \mathbb{E}_{\mathbf{C}_i, S^{est}}\bigg(\big(\theta(\mathbf{C}_i, \Pi) - \widetilde{\theta}(\mathbf{C}_i, S^{est}, \Pi)\big)^T\Sigma^{-1}\big(\theta(\mathbf{C}_i, \Pi) - \widetilde{\theta}(\mathbf{C}_i, S^{est}, \Pi)\big)\bigg) = \\
    & -\mathbb{E}_{\mathbf{C}_i}\big(\theta(\mathbf{C}_i, \Pi)^T\Sigma^{-1}\theta(\mathbf{C}_i, \Pi)\big) + \mathbb{E}(\operatorname{tr}(I))_{2 \times 2}.
\end{align}

The second equality follows after taking into account the independence of the train and estimation data,  $cov(A, B) = 0$. The final equality is based on the fact that $\theta(\mathbf{C}_i, \Pi) = \mathbb{E}(\theta(\mathbf{C}_i)|\mathbf{C}_i \in \ell(c, \Pi))$, $\mathbb{E}\big(\widetilde{\theta}(\mathbf{C}_i, S^{est}, \Pi)\big) = \theta(\mathbf{C}_i, \Pi)$,  and:
\begin{align*}
  \mathbb{E}_{\mathbf{C}_i, S^{est}}&\bigg(\big(\theta(\mathbf{C}_i, \Pi) - \widetilde{\theta}(\mathbf{C}_i, S^{est}, \Pi)\big)^T\Sigma^{-1}\big(\theta(\mathbf{C}_i, \Pi) - \widetilde{\theta}(\mathbf{C}_i, S^{est}, \Pi)\big)\bigg) = \\
  &\operatorname{tr}\bigg(\Sigma^{-1} \mathbb{E}\big(\theta(\mathbf{C}_i, \Pi) - \widetilde{\theta}(\mathbf{C}_i, S^{est}, \Pi)\big)^T\big(\theta(\mathbf{C}_i, \Pi) - \widetilde{\theta}(\mathbf{C}_i, S^{est}, \Pi)\big)\bigg) = \\
  & \operatorname{tr}(\Sigma^{-1}\Sigma) = \operatorname{tr}(I)_{2\times2},
\end{align*}
where $tr(I)_{2\times2}$ is the trace of a $2 \times 2$ identity matrix. Since $\mathbb{E}(tr(I))_{2 \times 2}$ does not depend on the parameter of interest, we can disregard it. Hence, the optimal parameter maximizes the unbiased estimator of the negative mean squared error:
\begin{align}\label{eq_loss_theta}
    \hat{\theta}(\mathbf{C}_i, S^{est}, \Pi) = \arg\max_{\widetilde{\theta}} \frac{1}{N^{tr}} \sum_{\ell} N^{tr}_{\ell}\widetilde{\theta}(\mathbf{C}_i,  \Pi)^{T}\widehat{\Sigma}^{-1}\widetilde{\theta}(\mathbf{C}_i, \Pi),
\end{align}
where the covariance matrix can be estimated as $\hat{\Sigma} =  \hat{\Sigma}\big(\tilde{\theta}(\mathbf{C}_i, S^{tr}, \Pi)|N^{est}\big)$. In this study, training and estimation samples have an equal number of observations, $N^{tr} = N^{est}$.

\section{Consistency of Component Weights }\label{app_brillinger}

\begin{proof}
We adopt the approach of \cite{naik2000partial} in which 
$$\mathbf{b}^\star = (\Sigma_{\mathbf{X}\mathbf{X}})^{-1}\sigma_{\mathbf{X}\mathbf{y}},$$
where $\Sigma_{\mathbf{X}\mathbf{X}}$ is the covariance matrix of the predictors and $\sigma_{\mathbf{X}\mathbf{y}}$ is the covariance vector between the predictors and the response. We define the matrix $R$ as follows:

$$R = (\sigma_{\mathbf{X}\mathbf{y}}, \Sigma_{\mathbf{X}\mathbf{X}}\sigma_{\mathbf{X}\mathbf{y}}, \dots, \Sigma_{\mathbf{X}\mathbf{X}}^{q-1}\sigma_{\mathbf{X}\mathbf{y}}),$$
where $q \leq p$ is a positive integer.

Under the assumption that $S_{xx}$ approaches $\Sigma_{\mathbf{X}\mathbf{X}}$ and $s_{xy}$ approaches $\sigma_{\mathbf{X}\mathbf{y}}$ as the sample size $N$ approaches infinity, we have:

\begin{align*}
    \hat{\mathbf{b}} \rightarrow R(R^T\Sigma_{\mathbf{X}\mathbf{X}}R)^{-1}\Sigma_{\mathbf{X}\mathbf{X}}\mathbf{b}^\star \ \text{ in probability when} \ N \rightarrow \infty. 
\end{align*}
The assumptions $q = M$ and $\sigma_{\mathbf{X}\mathbf{y}} = \sum_{j=1}^M\gamma_jv_j$ imply that $\mathbf{b}^\star$ is contained in the space spanned by $R$, and that $R$ is invertible. Consequently, by using the inverse properties,


\begin{align*}
    \hat{\mathbf{b}} \rightarrow R  R^{-1} \Sigma_{\mathbf{X}\mathbf{X}}^{-1} (R^T)^{-1} R^{T} \Sigma_{\mathbf{X}\mathbf{X}} \mathbf{b}^\star = \mathbf{b}^\star
\end{align*}

Hence, $\hat{\mathbf{b}} \rightarrow \mathbf{b}^\star$. \cite{brillinger2012generalized} in Sections 3 and 4 shows that $\sigma_{\mathbf{Xy}} = \xi \Sigma_{\mathbf{X}\mathbf{X}}\mathbf{b}$, where $\xi = \mathbf{cov\big(g(b_0 + \mathbf{X}\mathbf{b}), b_0 + \mathbf{X}\mathbf{b})\big)}/var\big(\mathbf{b_0 + \mathbf{X}\mathbf{b}}\big)$. Therefore, the proof is complete by noting that $\hat{\mathbf{b}} \rightarrow \Sigma_{\mathbf{X}\mathbf{X}}^{-1}\sigma_{\mathbf{X}\mathbf{y}} = \xi \mathbf{b}$. 

\end{proof}

\section{Asymptotic Properties of Policy Effects}\label{app_main_asymp}

The large sample theory of causal forests is largely based on the works of \cite{wager2018estimation} and \cite{wager2015adaptive}, \cite{nareklishvili2022adaptive} for partially identified policy effects, and \cite{li2020asymptotic} under network effects. It is important to note that the outcome variable $y_i$ in this context may be a vector-valued variable, in which case most of the operations and results apply coordinate-wise. 

\subsection{Asymptotic Normality}\label{app_hoeff}

Given a collection of ${\ell}_{n=1}^{|\Pi|}$ terminal nodes that form a partition of the component space $\mathbf{C}$, we define the prediction of a tree at a generic test point $c$ as:

\begin{align}
     \mathcal{T} = \mathcal{T}(c, \xi, A_i, \dots, A_N) = \sum_{n=1}^{|\Pi|} 1(c\in \ell_n) \frac{1}{N_{\ell_n}}\sum_{i:\mathbf{C}_i\in \ell_n}y_{i}.
\end{align}
$\xi$ is an external source of randomization, to allow for the randomized split selection procedures. $1(c \in \ell_n)$ is an indicator function and equals one if a point $c \in \ell_n$, and zero otherwise. $N_{\ell_n}$ denotes the number of observations in a terminal node $\ell_n$. 

A tree $\mathcal{T}(c, \xi, A_1, \dots, A_N)$ represents a prediction at a point $c$ based on data ${A}_{i = 1}^{N} = \{y_i, \mathbf{C}_i\}_{i = 1}^{N}$ and a randomization parameter $\xi$. As described in \cite{lewis2000introduction} and \cite{kingsford2008decision}, trees are a popular choice for classification and regression tasks due to their interpretability, ease of implementation, and robustness to outliers and missing data. However, trees also have a high variance and are prone to overfitting, which makes it difficult to determine the optimal tree structure. To address these issues, \cite{breiman2001random} introduced the random forest algorithm.

Let $s < N$ be a subset of size $s$ from a population $i = {1, \dots, N}$, where $s = N^{\beta}$ and $\beta$ is sufficiently close to 1 (\citealp{wager2018estimation}). Following the work of \cite{breiman2001random} and \cite{wager2018estimation}, we define the random forest estimator as the average of the tree estimators aggregated over all possible size-$s$ subsamples of the training data, marginalized over the auxiliary noise $\xi$. Specifically, the prediction of the random forest estimator at a particular test data point $c$ is defined as:

\begin{align}\label{eq_rf}
    \mathcal{F}(c, A_1, \dots, A_N) = \frac{1}{\big(\begin{smallmatrix} N \\ s \end{smallmatrix}\big)}\sum_{1 \leq i_1 \leq  \dots \leq i_s \leq N} \mathbb{E}_{\xi}\mathcal{T}(c, \xi, A_{i_1}, \dots, A_{i_s}),
\end{align}
where $i_1, \dots, i_s$ are the size-s subsamples of the population $\{i = 1, \dots, N\}$. In practice, we estimate such a random forest by Monte Carlo averaging:
\begin{align}
 \mathcal{F}(c, A_1, \dots, A_N) \approx \frac{1}{B}\sum_{b=1}^{B}\mathcal{T}(c, \xi^*,  A^*_{1}, \dots A^*_{N})
\end{align}
where $\{A^*_{1}, \dots A^*_{N}\}$ is drawn without replacement from  $\{A_{1}, \dots A_{N}\}$. $\xi^*$ is an auxiliary noise in a given sample and $B$ is the number of sub-samples. $\mathcal{F}(c, A_1, \dots, A_N)$ is a $1 \times M$ vector. Therefore, most of the arithmetic operations in this section are defined coordinate-wise in $\mathbb{R}^M$.

A random forest estimator can be represented as a U-statistic (\citealp{hoeffding1961strong, korolyuk2013theory}). A common approach to studying the large sample properties of random forests is to derive the lower bound of its Höeffding decomposition. Höeffding decomposition (also known as the Hajek projection) in a univariate setting is described by  \cite{hajek1968asymptotic}. Specifically, consider a vector-valued function $T \in \mathbb{R}^M$ which is measurable and permutation symmetric, that is,  $T(\pi c) = T(c)$ for all $\pi \in \Pi$ (a tree in this setting). Then the Hajek projection of this function is defined as:
\begin{align}\label{eq_hajek_pro}
    \dot{T} = \mathbb{E}(T) + \sum_{i=1}^N \big[\mathbb{E}(T|\mathbf{C}_i) - \mathbb{E}(T)\big] = \sum_{i= 1}^N\mathbb{E}(T|\mathbf{C}_i) - (N-1)\mathbb{E}(T). 
\end{align}

Intuitively, the Hajek projection in \eqref{eq_hajek_pro} represents a projection of $T$ onto the linear subspace of all random variables of the form $\sum_{i=1}^Ng_i(\mathbf{C}_i)$, where $g_i: \mathbb{R}^d \rightarrow \mathbb{R}$ are arbitrary measurable functions such that $\mathbb{E}(g_i^2(\mathbf{C}_i)) < \infty$ for $i = 1, \dots, N$. It is clear that the conditional expectation of the centered and symmetric component $\dot{T}$ in \eqref{eq_hajek_pro} is equal to the conditional expectation of $T$:

\begin{align}
    \mathbb{E}(\dot{T}|\mathbf{C}_i) &= \mathbb{E}(T|\mathbf{C}_i), \ \text{and} \\
    \mathbb{E}(\dot{T}) &= \mathbb{E}(T).\nonumber
\end{align}

Now consider the random forest estimator, $\mathcal{F}(c, A_1, \dots, A_N) \in \mathbb{R}^M$, and let the corresponding vector of means be $\mu$. Moreover, let $\dot{\mathcal{F}}(c, A_1, \dots, A_N)$ and $\Sigma$ denote the Hajek projection of the random forest estimator, and the covariance matrix of the Hajek projection, respectively. Assume also that the trees in $\dot{\mathcal{F}}(c, A_1, \dots, A_N)$ are symmetric and the observations are $i.i.d$. 
Under the given assumptions,
\cite{wager2018estimation} show that the random forest estimator is consistent and asymptotically normal. They generalize the properties to a single parameter of interest. Additionally, \cite{nareklishvili2022adaptive} in Theorem 6.3 shows that the causal forest estimator is asymptotically normally distributed even for multiple (possibly correlated) parameters.

\begin{figure}[H]
	\centering
	\subfloat[N = 500]{\includegraphics[width=0.5\linewidth]{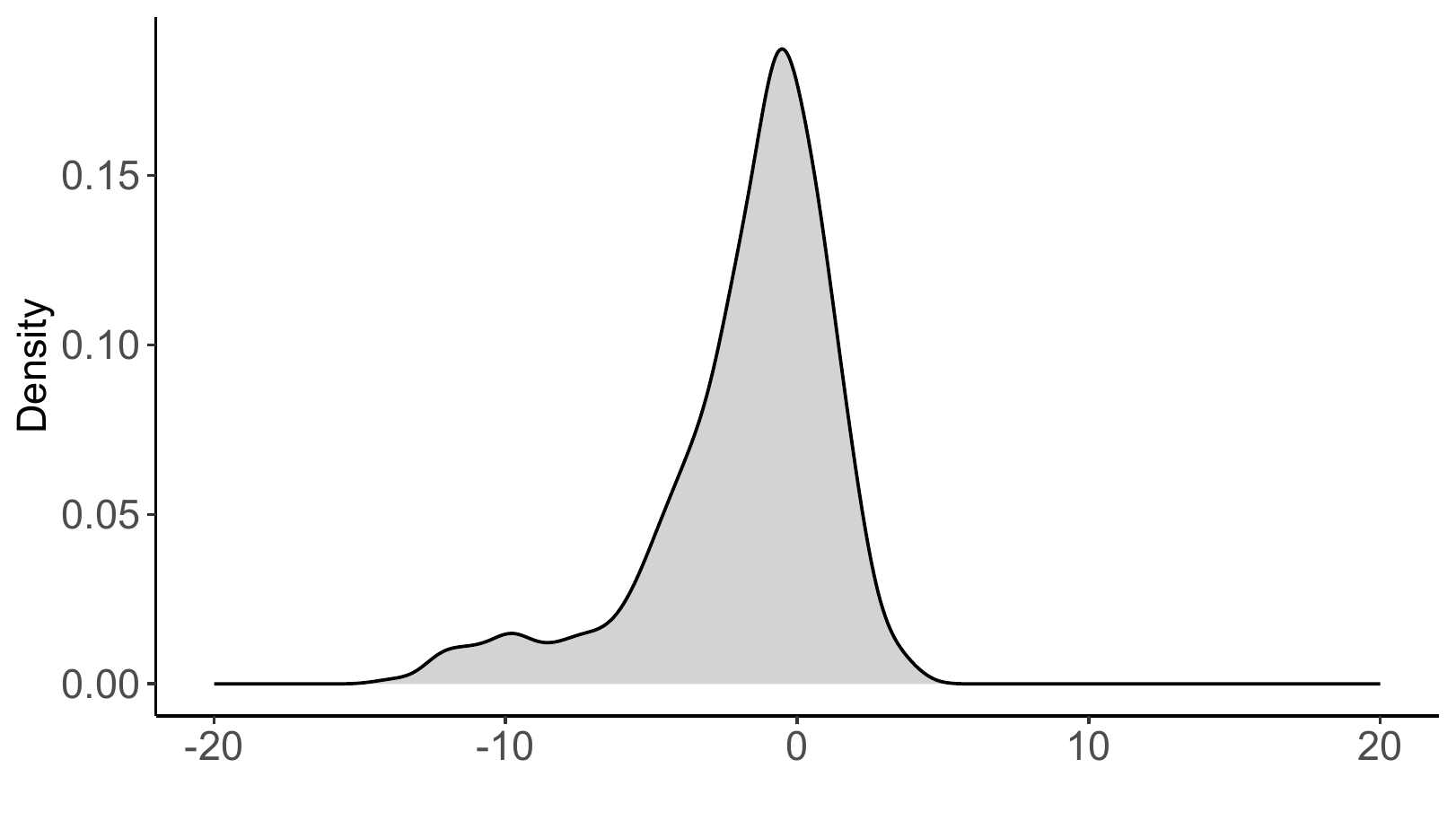}} 
   \subfloat[N = 1000] {\includegraphics[width=0.5\linewidth]{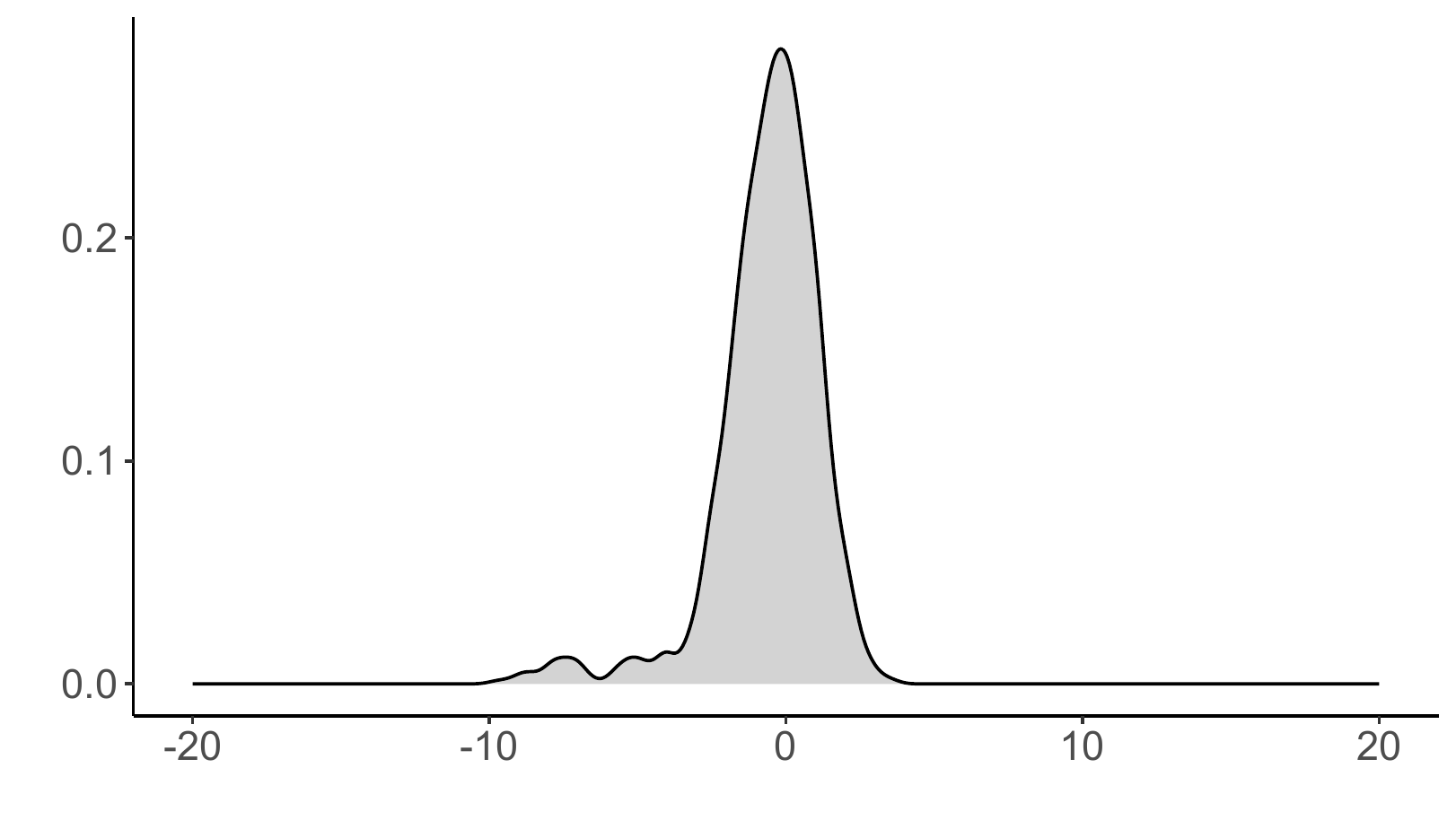}} \quad
	\subfloat[N = 10,000]{\includegraphics[width=0.5\linewidth]{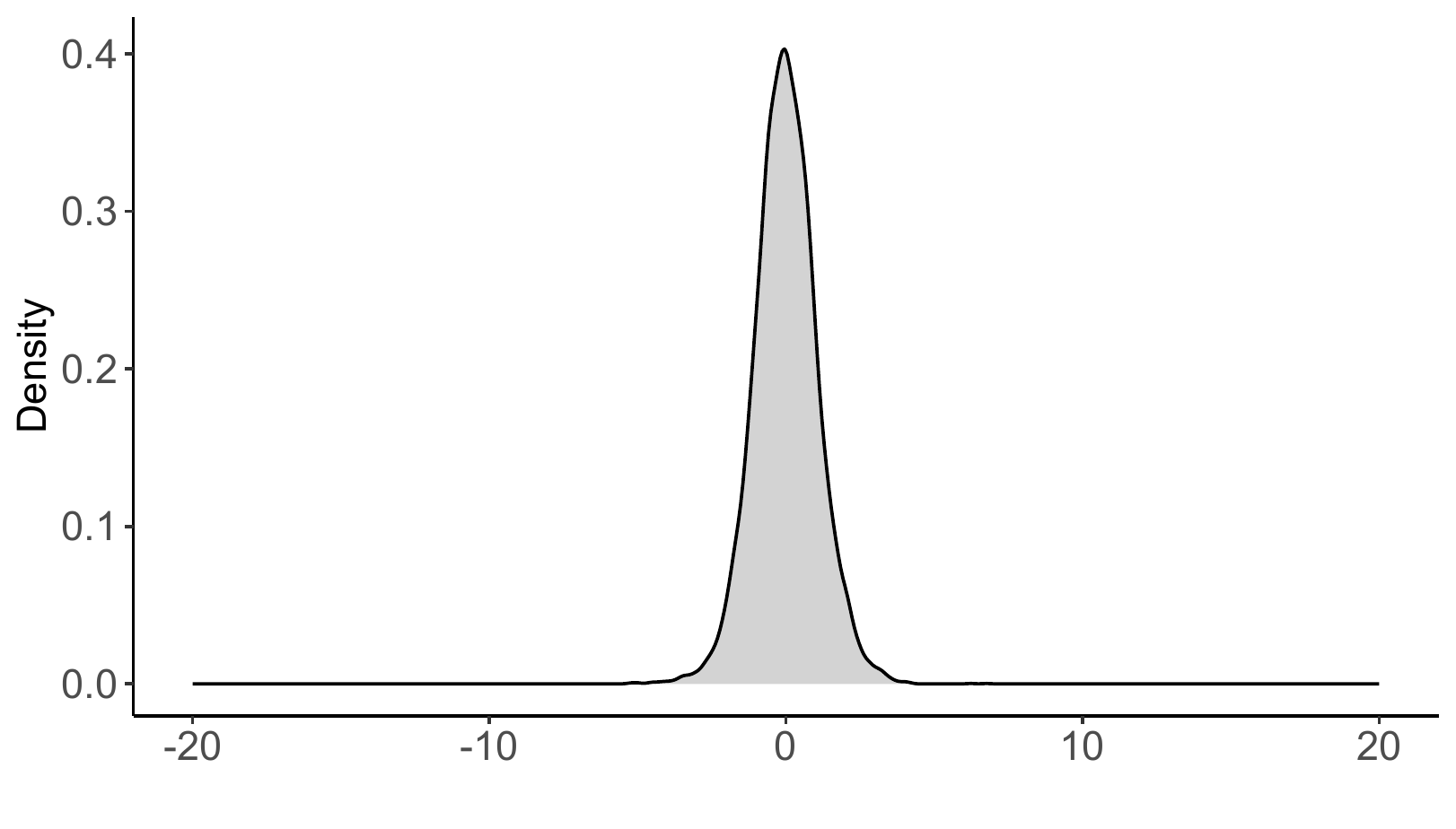}}
    \subfloat[N = 100,000] {\includegraphics[width=0.5\linewidth]{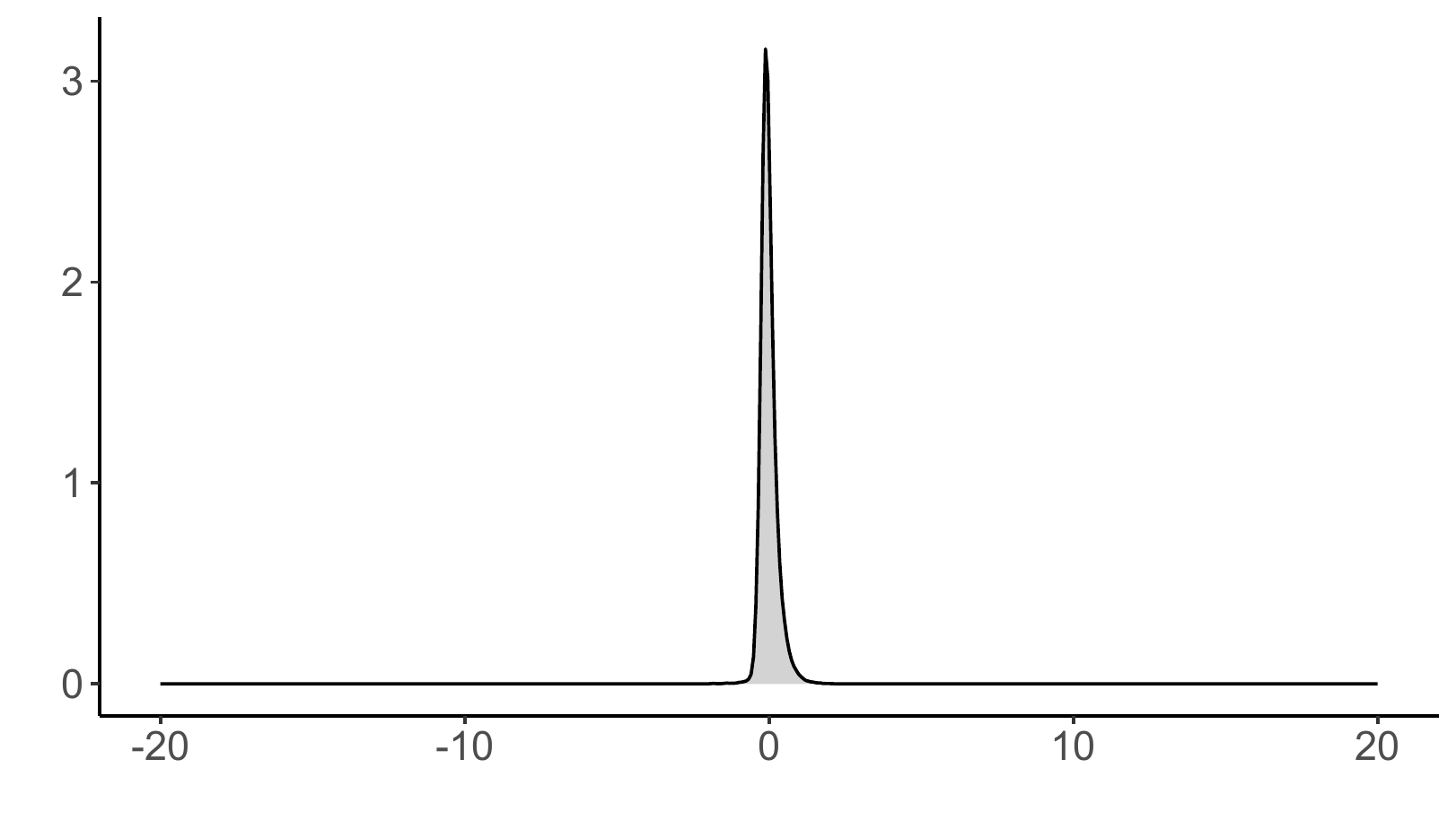}} \quad 
	\caption{The difference between the densities of estimated and simulated policy effects.  }\label{fig_theory}
\end{figure}

Figure \ref{fig_theory} presents a validation analysis comparing the density of estimated policy effects to simulated policy effects across various sample sizes. The findings indicate that when $N \rightarrow \infty$, $\mathcal{F}(c, A_1, \dots, A_N) - \dot{\mathcal{F}}(c, A_1, \dots, A_N) \xrightarrow{p} 0$.  According to the theoretical framework illustrated in Figure \ref{fig_theory}, it can be inferred that the discrepancy between the estimated and simulated policy effects exhibits asymptotic normality. 

\subsection{Inference}\label{app_sub_inference}
We quantify the uncertainty of policy effects by using the jackknife variance estimator (as described in \citealp{wager2018estimation}). Let $g = 1, \dots, G$ be the $g$-th bootstrapped sample. We use a tree $\Pi_g$ and the corresponding estimation sample $S_g^{est}$ to obtain $\hat{\theta}_g(c, S_g^{est}, \Pi_g)$ at a generic test point $c$.  Next, the average of the individual tree estimates are given as:

$$\hat{\theta}\big(c, \{S_g^{est}\}_{g=1}^B\}, \{\Pi_g\}_{g=1}^G\big) = \frac{1}{G}\sum_{g = 1}^G\hat{\theta}_g(c, S_g^{est}, \Pi_g).$$ 

We define $N_{ig}$ as the number of times an observation $i$ appears in the $g$-th bootstrapped sample, either in the training sample $S^{tr}$ or the estimation sample $S^{est}$. The following variance estimator can be used to construct valid confidence intervals::

\begin{align}
    &Var\big[\hat{\theta}\big(c, \{S_g^{est}\}_{g=1}^G\}, \{\Pi_g\}_{g=1}^G\big) \big] =  \sum_{i = 1}^N \Delta_g - \frac{N}{G^2}\sum_{g=1}^G\big[ {\theta}_g(c, S_g^{est}, \Pi_g) -  \hat{\theta}\big(c, \{S_g^{est}\}_{g=1}^G\}\big)     \big]^2,    \\
    & \text{where} \ \Delta_g = \bigg [ \frac{\sum_{g = 1}^G(N_{ig}-1)\big[\hat{\theta}_g(c, S_G^{est}, \Pi_g) -  \hat{\theta}\big(c, \{S_g^{est}\}_{g=1}^G\}, \{\Pi_g\}_{g=1}^G\big)\big]}{G}\bigg]. \nonumber
\end{align}

\section{Contribution of Individual Characteristics} \label{app_contr}

\begin{table}[H] \centering 
  \caption{Target components of Forest-PLS based on the simulated experiment in \eqref{eq_rct_v2}. The optimal number of components is chosen by a five-fold cross-validation approach. Subsequently, we run a linear regression to assess the influence of the independent variables on each target component. The reported results represent the average findings obtained from fifty different replications of the experiment. } 
  \label{tab_tc} 
  \resizebox{\textwidth}{!}{%
\begin{tabular}{@{\extracolsep{5pt}}lcccccccc} 
\\[-1.8ex]\hline 
\hline \\[-1.8ex] 
 \multicolumn{6}{c}{\textit{Dependent variable:}} & \\ 
\cline{2-7} \\
 &  \multicolumn{1}{c}{$N = 100$}  & 
 \multicolumn{3}{c}{$N = 1000$} &  \multicolumn{3}{c}{$N = 5000$} & \\
 \cline{2-3} \cline{4-5}  \cline{6-7}
\\[-1.8ex] & $C_1$ & $C_2$ & $C_1$ & $C_2$ & $C_1$ & $C_2$ & \\ 
\hline \\[-1.8ex] 
 $X_1$ & 0.748$^{***}$ & 0.103$^{***}$ & 0.729$^{***}$ & 0.104$^{***}$ & 0.708$^{***}$ & 0.035$^{***}$ \\ 
  & (0.000) & (0.000) & (0.000) & (0.000) & (0.000) & (0.000) \\ 
  & & & & & & \\ 
  $X_2$ & 0.730$^{***}$ & 0.209$^{***}$ & 0.729$^{***}$ & 0.100$^{***}$ & 0.708$^{***}$ & 0.040$^{***}$ \\ 
  & (0.000) & (0.000) & (0.000) & (0.000) & (0.000) & (0.000) \\ 
  & & & & & & \\ 
  $X_3$ & $-$0.096$^{***}$ & 0.795$^{***}$ & $-$0.039$^{***}$ & 0.585$^{***}$ & 0.015$^{***}$ & $-$0.432$^{***}$ \\ 
  & (0.000) & (0.000) & (0.000) & (0.000) & (0.000) & (0.000) \\ 
  & & & & & & \\ 
  $X_4$ & 0.082$^{***}$ & $-$0.668$^{***}$ & $-$0.057$^{***}$ & 0.819$^{***}$ & 0.024$^{***}$ & $-$0.907$^{***}$ \\ 
  & (0.000) & (0.000) & (0.000) & (0.000) & (0.000) & (0.000) \\ 
  & & & & & & \\ 
 Constant & 0.306$^{***}$ & $-$1.892$^{***}$ & 0.085$^{***}$ & $-$1.196$^{***}$ & $-$0.040$^{***}$ & 0.858$^{***}$ \\ 
  & (0.000) & (0.000) & (0.000) & (0.000) & (0.000) & (0.000) \\ 
  & & & & & & \\ 
\hline \\[-1.8ex] 
Observations & 100 & 10,000 & 1,000 & 10,000 & 10,000 & 10,000 \\ 
R$^{2}$ & 1.000 & 1.000 & 1.000 & 1.000 & 1.000 & 1.000 \\ 
Adjusted R$^{2}$ & 1.000 & 1.000 & 1.000 & 1.000 & 1.000 & 1.000 \\ 
\hline 
\hline \\[-1.8ex] 
\textit{Note:}  & \multicolumn{6}{r}{$^{*}$p$<$0.1; $^{**}$p$<$0.05; $^{***}$p$<$0.01}  \\
\end{tabular} }
\end{table}

\begin{table}[H]
\centering
\caption{Variable importance for policy effects estimated by the Forest-PLS and causal forest methods, respectively. To quantitatively assess the magnitude of individual characteristics contributing to policy effects, we employ ordinary regression trees with the policy effect as the outcome variable. 
Both methods consist of 1000 trees. The simulation is based on the setup suggested in \eqref{eq_rct_v2}. The reported results represent the average findings obtained from fifty different replications of the experiment. }
\label{tab_varimp}
\begin{tabular}{@{\extracolsep{6pt}} cccccccccc }
\\[-1.8ex]\hline
& \multicolumn{2}{l}{N = 100} &  \multicolumn{3}{l}{N = 1000} & \multicolumn{1}{l}{N = 5000}  \\
\cmidrule(r{0.9pc}){2-3} \cmidrule(r{0.9pc}){4-5} \cmidrule(r{0.9pc}){6-7} 
& Forest-PLS &  CF & Forest-PLS &  CF  & Forest-PLS &  CF\\
\hline \\[-1.8ex]
$X_1$ &  $0.266$ & $0.192$& $0.242$ & $0.265$ & $0.078$ & $0.157$ \\
$X_2$ &  $0.228$ & $0.252$ & $0.258$ & $0.124$ & $0.097$ & $0.126$ \\
$X_3$ & $0.291$ & $0.281$ & $0.323$ & $0.491$ &  $0.198$ & $0.427$\\
$X_4$ & $0.215$ & $0.276$ & $0.177$ & $0.120$ & $0.627$ & $0.290$\\
\hline \\[-1.8ex]
\end{tabular}
\end{table}

\begin{table}[H] \centering 
  \caption{The coefficients estimated by the LASSO regression. LASSO drops $X_{3}$ and $X_{4}$ that are relevant for explaining policy effect heterogeneity. The optimal value of the shrinkage parameter $\lambda$ is 2.605 and is chosen by cross-validation. } 
  \label{tab_lasso} 
\begin{tabular}{@{\extracolsep{5pt}} cc} 
\\[-1.8ex]\hline 
\hline \\[-1.8ex] 
Intercept & $1.121$ \\
$X_{1}$ & $97.264$ \\
$X_{2}$ &$97.201$ \\ 
$X_{3}$ & $0.000$  \\ 
$X_{4}$ & $0.000$  \\ 
\hline \\[-1.8ex] 
\end{tabular} 
\end{table}

\section{Additional Simulated Experiments}\label{app_additional_sims}
\begin{figure}[H]
	\centering
	\subfloat[N = 100]{\includegraphics[width=0.5\linewidth]{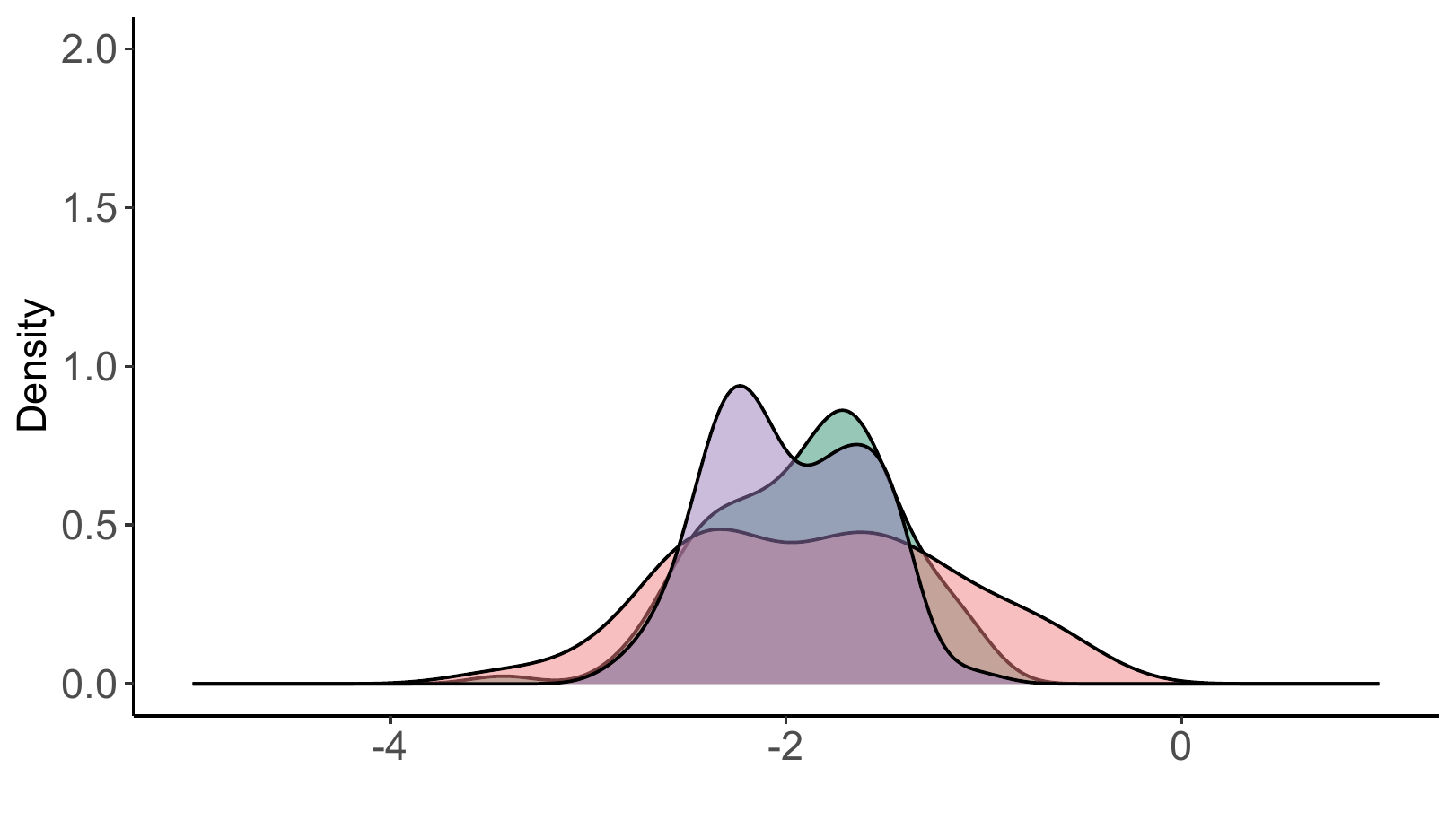}} 
   \subfloat[N = 500] {\includegraphics[width=0.5\linewidth]{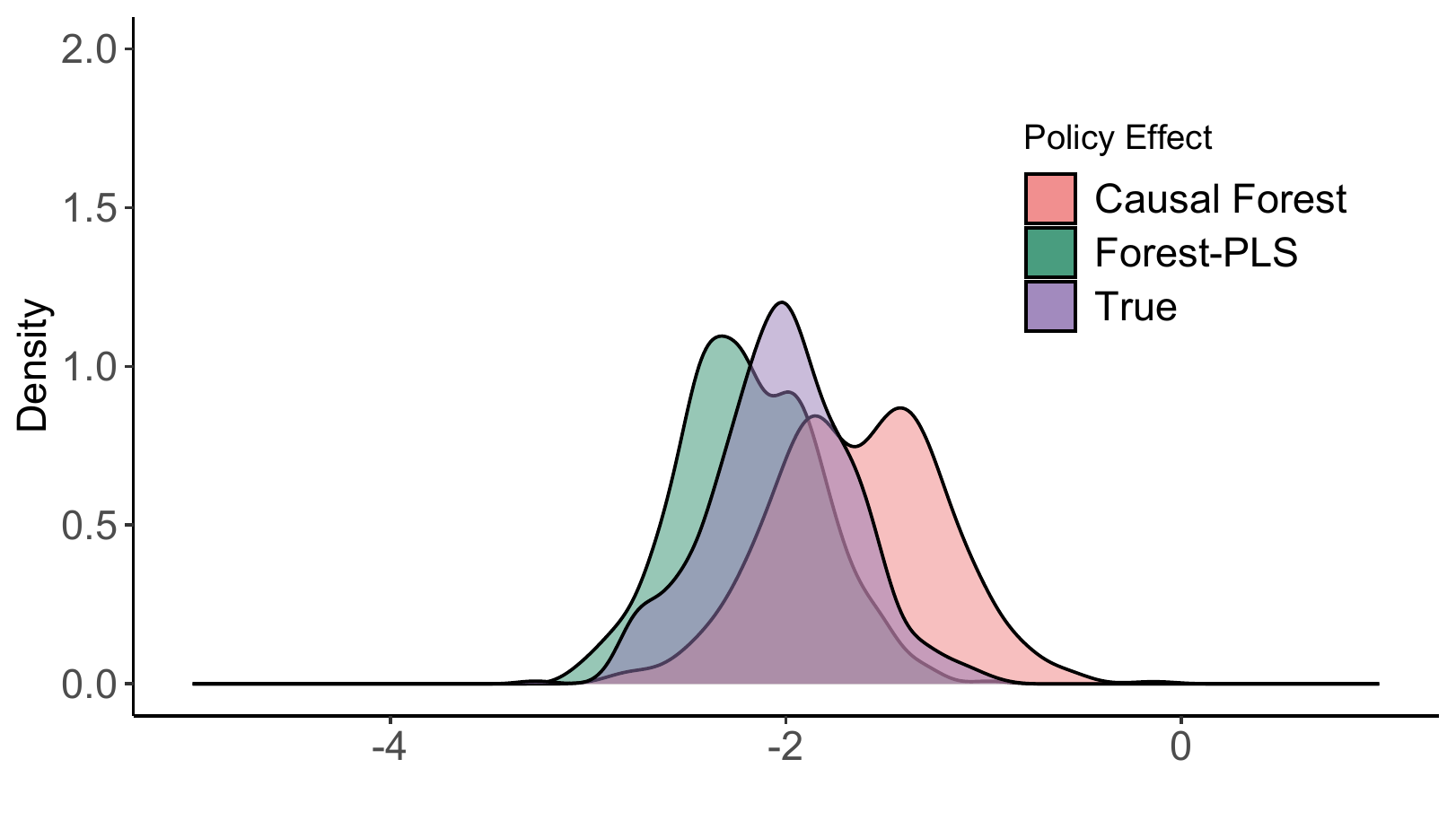}} \quad
	\subfloat[N = 1000]{\includegraphics[width=0.5\linewidth]{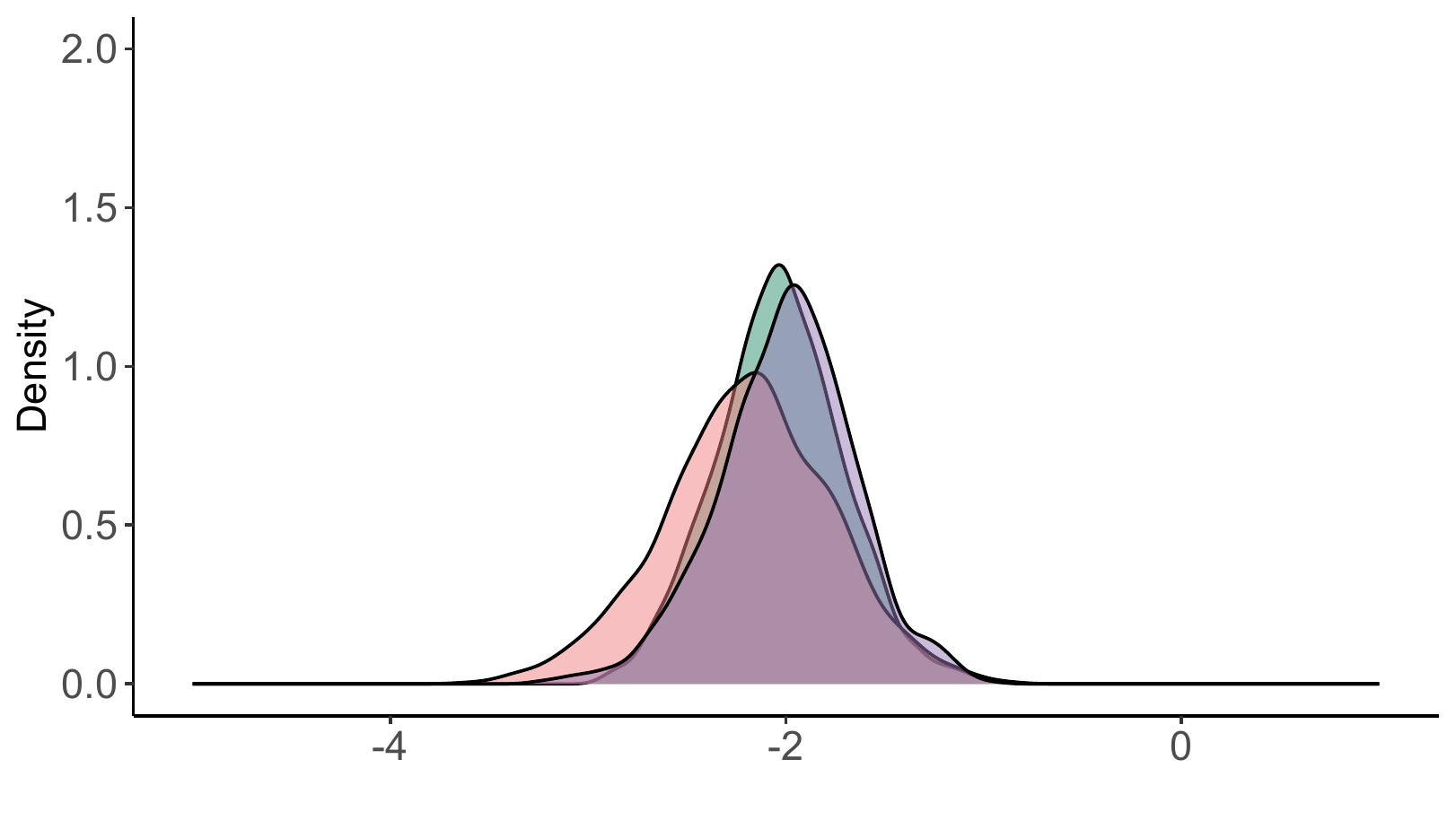}}
    \subfloat[N = 5000] {\includegraphics[width=0.5\linewidth]{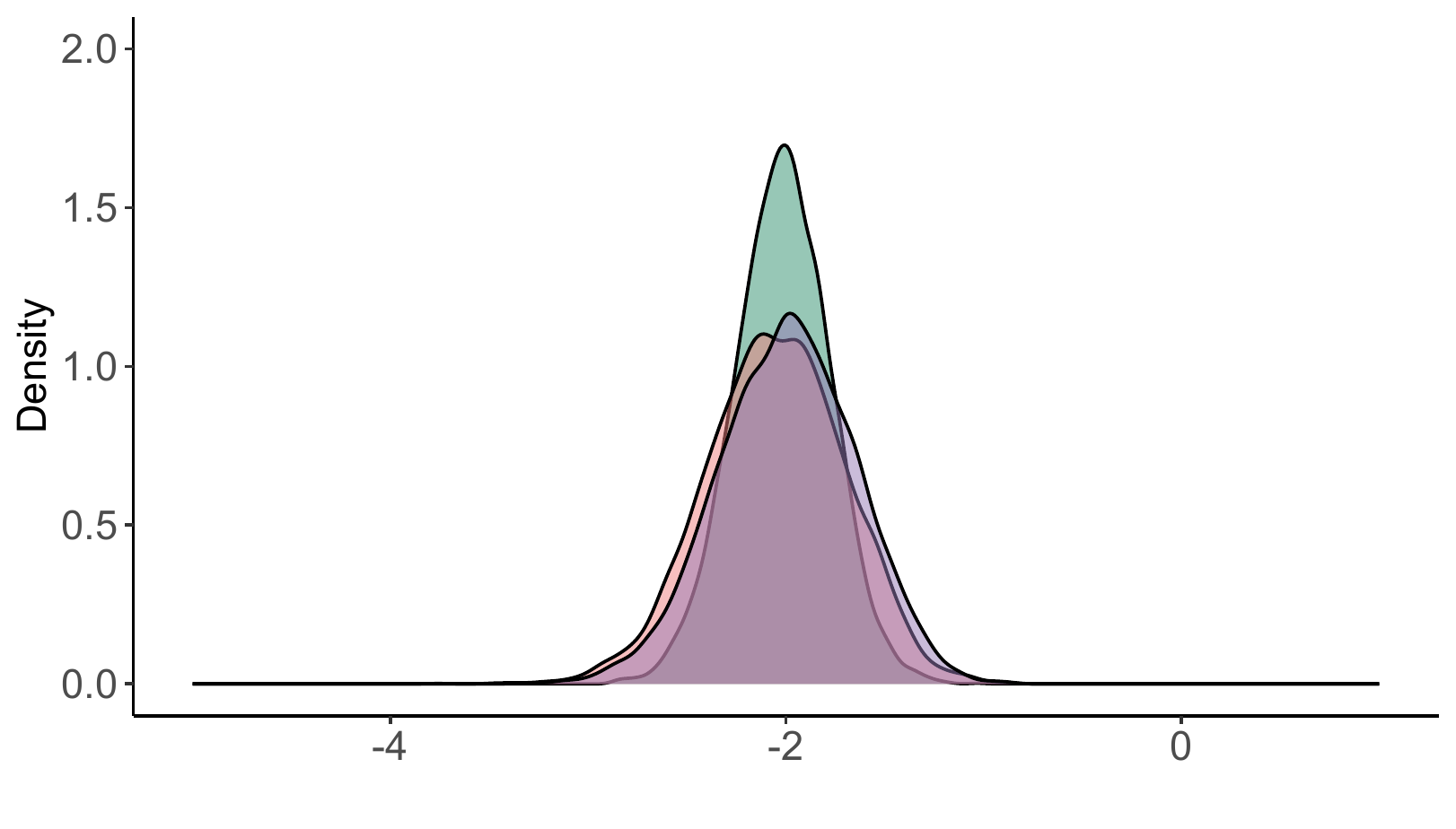}} \quad 
	\caption{Simulated experiments where the simulation design is the same as in \eqref{eq_rct_v2}, but the policy effect equals $X_{i3}\cdot X_{i1}$. In each experiment, the number of trees equals 1000. Forest-PLS consists of two optimal target components chosen by five-fold cross-validation. The reported distributions are the averages across fifty different replications of the experiment.   }\label{fig_group_heterogeneity_noconf}
\end{figure}

\begin{figure}[H]
	\centering
	\subfloat[N = 100]{\includegraphics[width=0.5\linewidth]{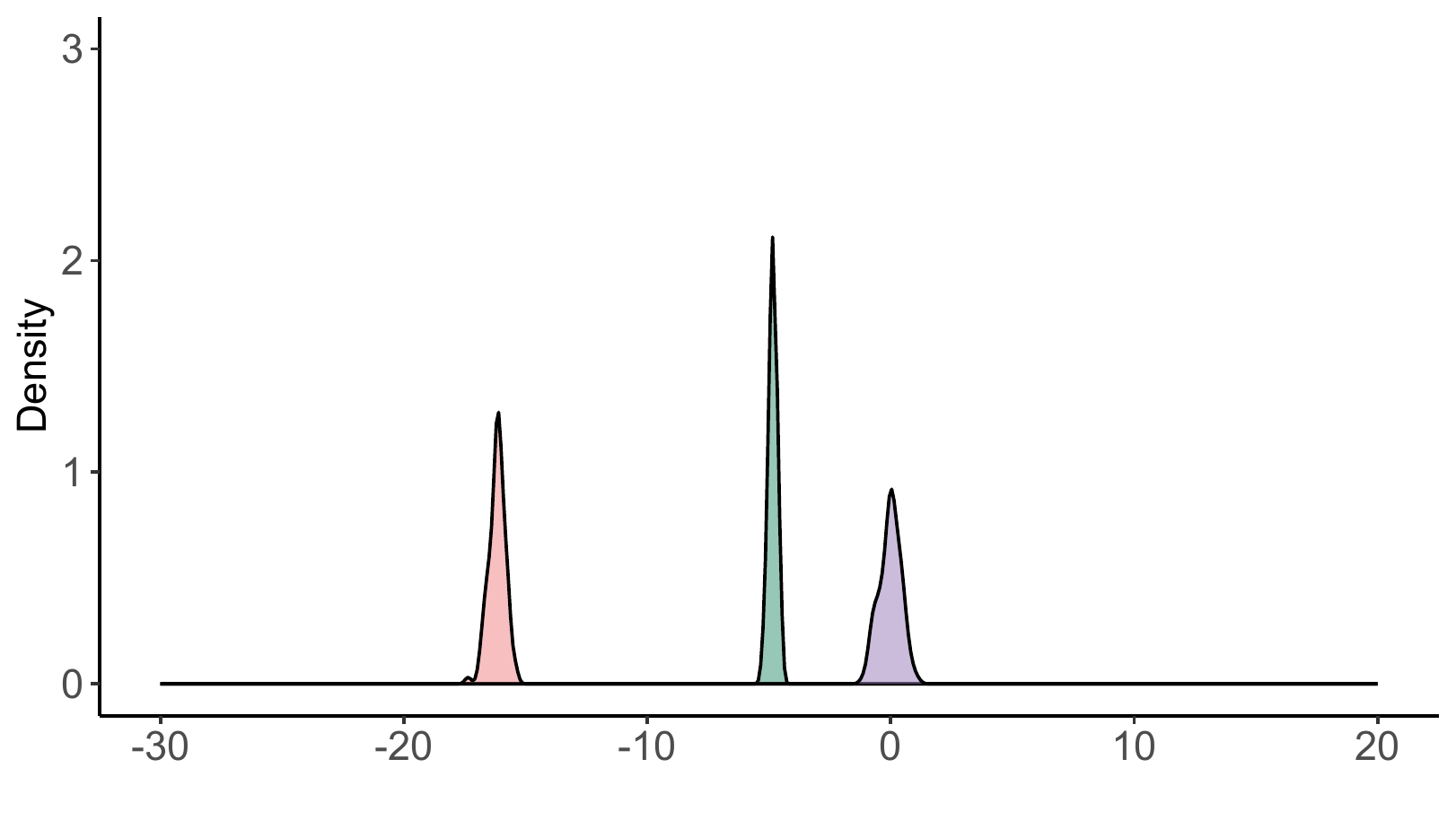}} 
   \subfloat[N = 500] {\includegraphics[width=0.5\linewidth]{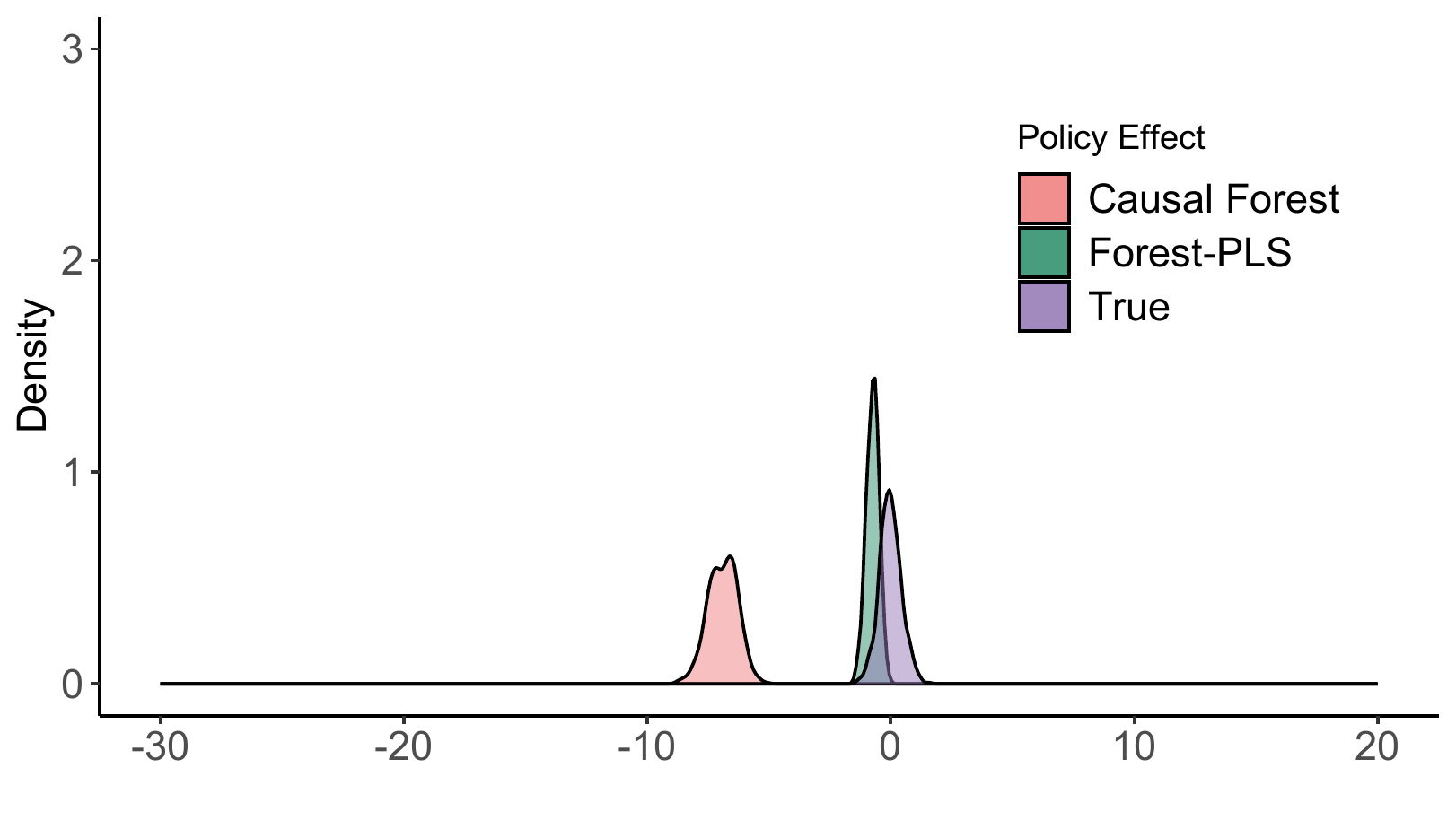}} \quad
	\subfloat[N = 1000]{\includegraphics[width=0.5\linewidth]{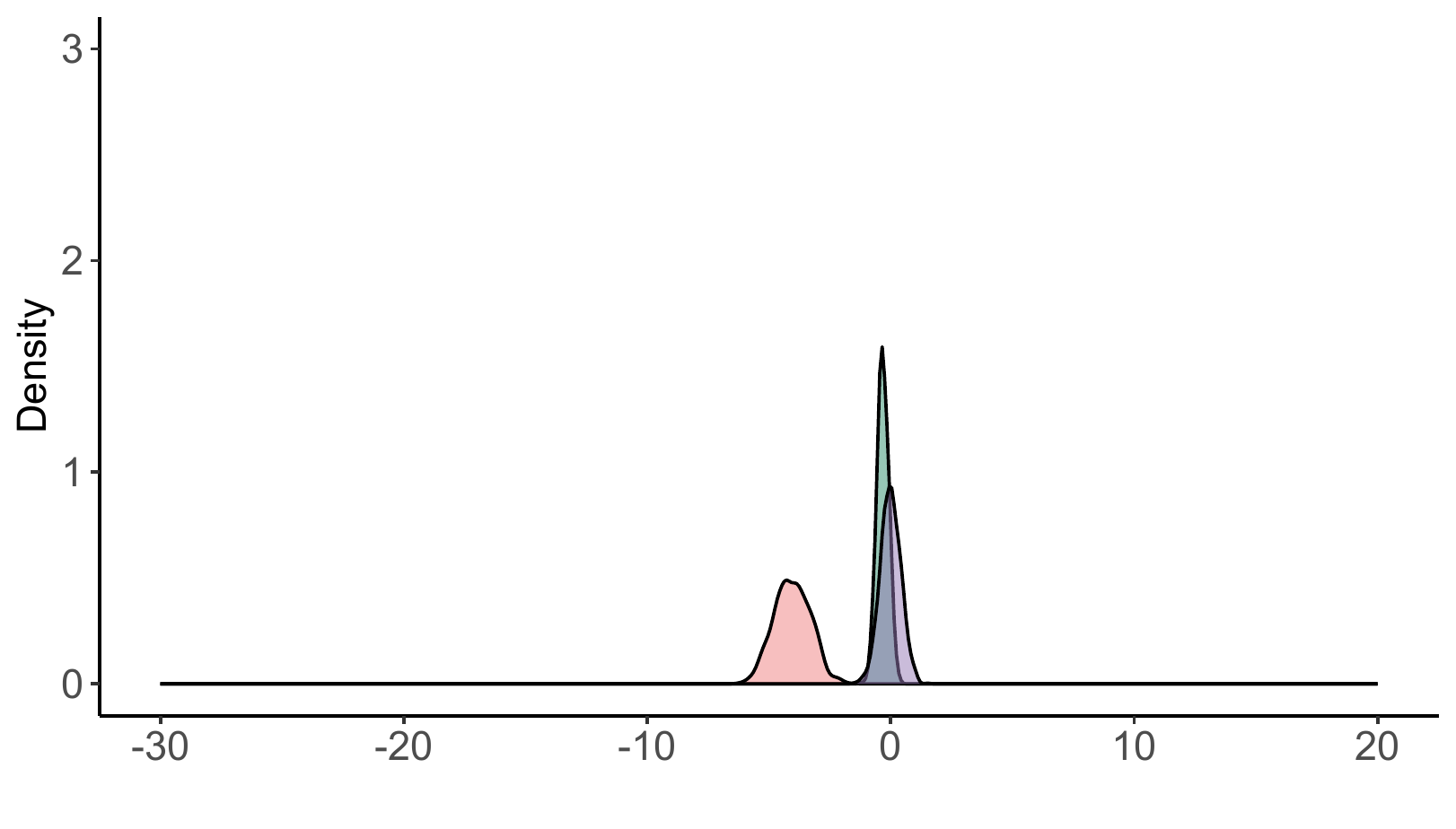}}
    \subfloat[N = 5000] {\includegraphics[width=0.5\linewidth]{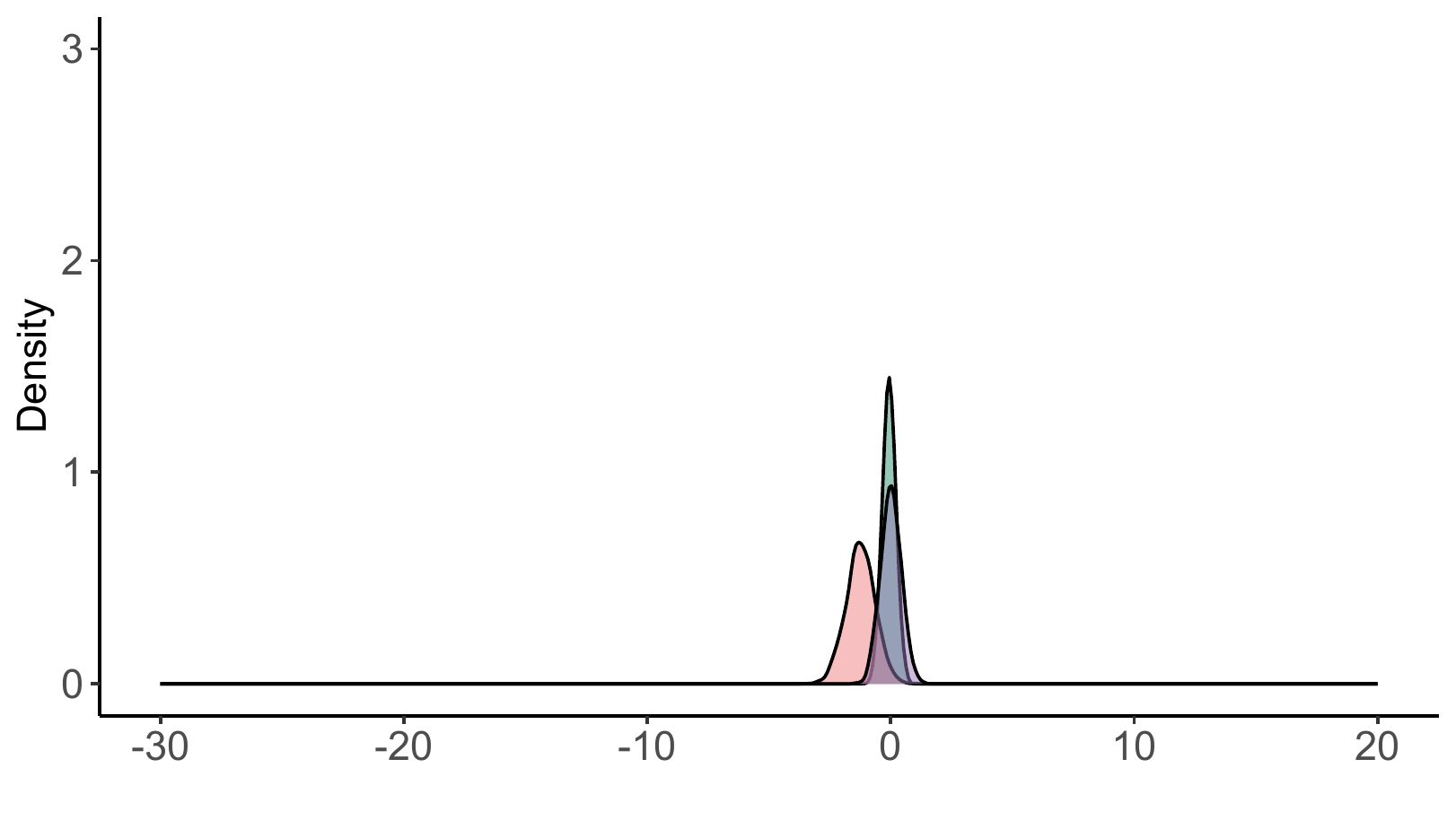}} \quad 
	\caption{Simulated experiments with  $D_i = X_{i1} - X_{i2} + 2\cdot X_{i3} + \mathcal{N}(0, 1)$ and $Y_i = 100\cdot X_{i1} + 100 \cdot X_{i2} - 100\cdot X_{i3} + 3\cdot D_i \cdot X_{i1} + \mathcal{N}(0, 1)$. In each experiment, the number of trees equals 1000. Forest-PLS consists of two optimal target components chosen by five-fold cross-validation. The reported distributions are the averages across fifty different replications of the experiment.   }\label{fig_group_heterogeneity_nbd}
\end{figure}

\section{Interpreting Target Components}\label{app_expl}

\begin{figure}[H]
\centering
\begin{tabular}{cc}
\includegraphics[width = \linewidth]{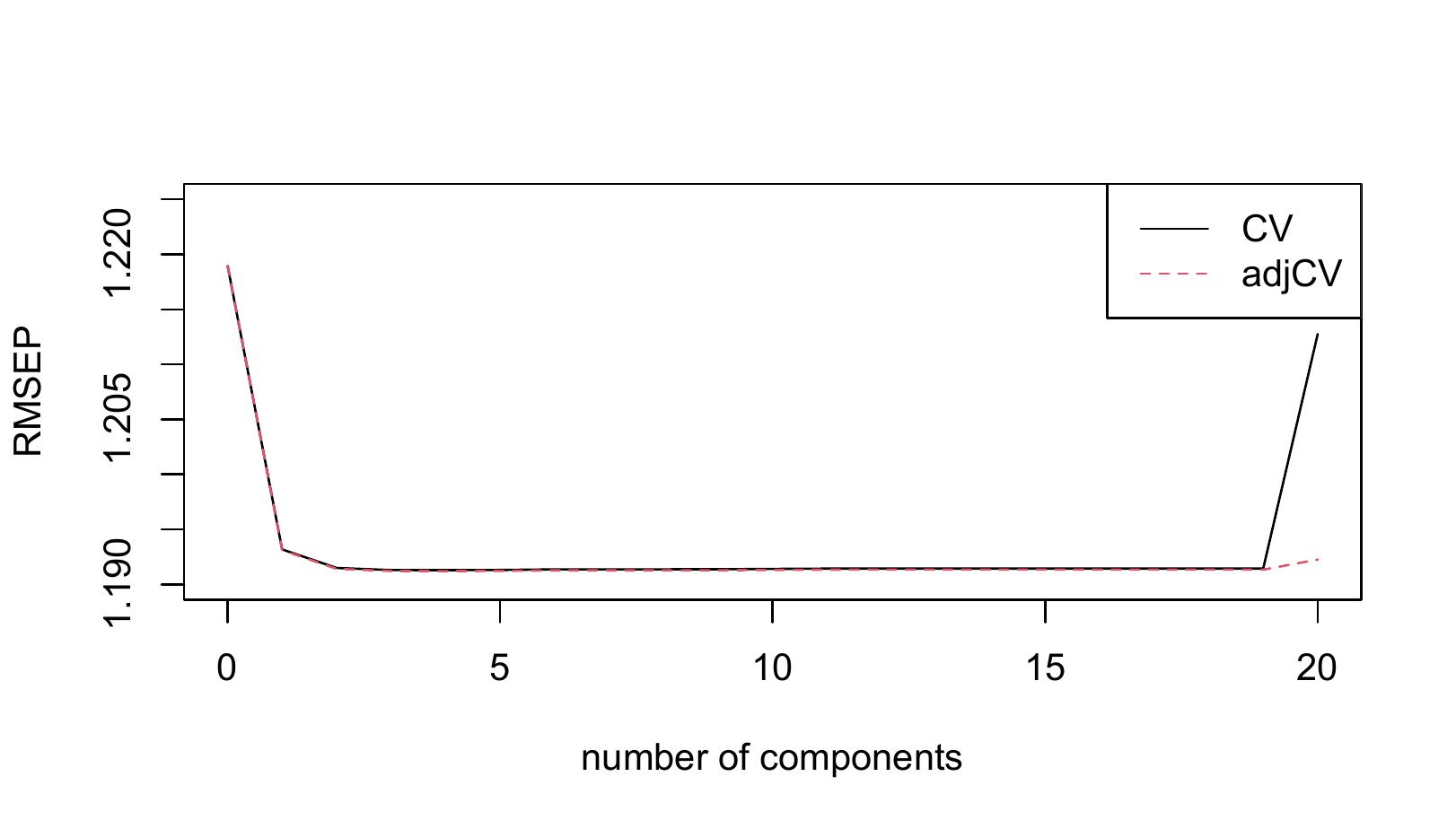} 
\end{tabular}
\caption{Root-mean-squared error (RMSE) based on five-fold cross-validation. }\label{fig_cv}
\end{figure}

Throughout the conducted experiments, we determine the number of optimal target components based on a five-fold cross-validation approach. The results, illustrated in Figure \ref{fig_cv}, demonstrate the convergence of the root-mean-squared-error (RMSEP) to a stable value of 1.191 as the number of components progressively increases to two or more. Consequently, we choose the minimum number of components that exhibit the lowest RMSEP. 

\begin{table}[H] \centering 
  \caption{Linear regression of each target component on independent characteristics. } 
  \label{app_table_comp} 
\small 
\resizebox{10cm}{!}{
\begin{tabular}{@{\extracolsep{5pt}}lcc} 
\\[-1.8ex]\hline 
\hline \\[-1.8ex] 
 & \multicolumn{2}{c}{\textit{Dependent variable:}} \\ 
\cline{2-3} 
\\[-1.8ex] & (Component 1) & (Component 2)\\ 
\hline \\[-1.8ex] 
 abdt & $-$0.001$^{***}$ & 0.004$^{***}$ \\ 
  & (0.000) & (0.000) \\ 
  female & 0.362$^{***}$ & 0.365$^{***}$ \\ 
  & (0.000) & (0.000) \\ 
  black & $-$1.270$^{***}$ & $-$0.955$^{***}$ \\ 
  & (0.000) & (0.000) \\ 
  hispanic & $-$0.901$^{***}$ & $-$0.631$^{***}$ \\ 
  & (0.000) & (0.000) \\ 
  othrace & $-$0.827$^{***}$ & $-$0.699$^{***}$ \\ 
  & (0.000) & (0.000) \\ 
  dep & 0.166$^{***}$ & 0.082$^{***}$ \\ 
  & (0.000) & (0.000) \\ 
  q1 & $-$0.467$^{***}$ & $-$2.588$^{***}$ \\ 
  & (0.000) & (0.000) \\ 
  q2 & $-$0.066$^{***}$ & $-$1.716$^{***}$ \\ 
  & (0.000) & (0.000) \\ 
  q3 & $-$0.516$^{***}$ & $-$1.334$^{***}$ \\ 
  & (0.000) & (0.000) \\ 
  q4 & $-$0.586$^{***}$ & $-$1.344$^{***}$ \\ 
  & (0.000) & (0.000) \\ 
  q5 & $-$0.847$^{***}$ & $-$0.602$^{***}$ \\ 
  & (0.000) & (0.000) \\ 
  recall & 1.473$^{***}$ & 0.724$^{***}$ \\ 
  & (0.000) & (0.000) \\ 
  agelt35 & $-$0.938$^{***}$ & $-$0.086$^{***}$ \\ 
  & (0.000) & (0.000) \\ 
  agegt54 & 1.238$^{***}$ & $-$0.361$^{***}$ \\ 
  & (0.000) & (0.000) \\ 
  durable & 0.190$^{***}$ & 0.018$^{***}$ \\ 
  & (0.000) & (0.000) \\ 
  nondurable & $-$0.724$^{***}$ & $-$0.989$^{***}$ \\ 
  & (0.000) & (0.000) \\ 
  lusd & $-$0.396$^{***}$ & $-$0.356$^{***}$ \\ 
  & (0.000) & (0.000) \\ 
  husd & 0.221$^{***}$ & $-$0.812$^{***}$ \\ 
  & (0.000) & (0.000) \\ 
  muld & 0.123$^{***}$ & 0.938$^{***}$ \\ 
  & (0.000) & (0.000) \\ 
  Constant & 10.317$^{***}$ & $-$39.629$^{***}$ \\ 
  & (0.000) & (0.000) \\ 
 \hline \\[-1.8ex] 
Observations & 13,913 & 13,913 \\ 
R$^{2}$ & 1.000 & 1.000 \\ 
Adjusted R$^{2}$ & 1.000 & 1.000 \\ 
Residual Std. Error (df = 13893) & 0.000 & 0.000 \\ 
\hline 
\hline \\[-1.8ex] 
\textit{Note:}  & \multicolumn{2}{r}{$^{*}$p$<$0.1; $^{**}$p$<$0.05; $^{***}$p$<$0.01} \\ 
\end{tabular} }
\end{table}

\begin{figure}[H]
	\centering
	\subfloat[Component 1]{\includegraphics[width=0.5\linewidth]{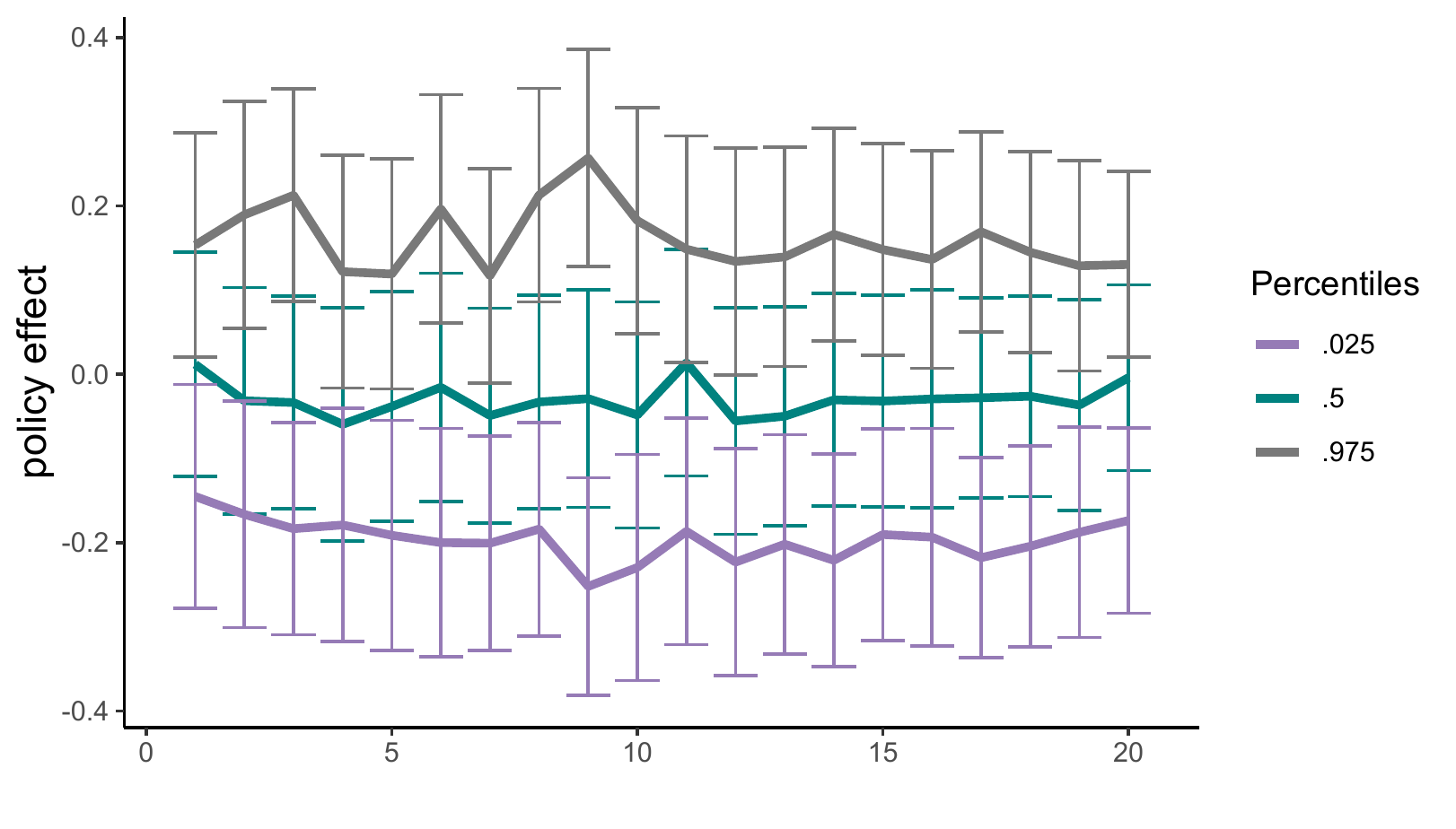}} 
   \subfloat[Component 2] {\includegraphics[width=0.5\linewidth]{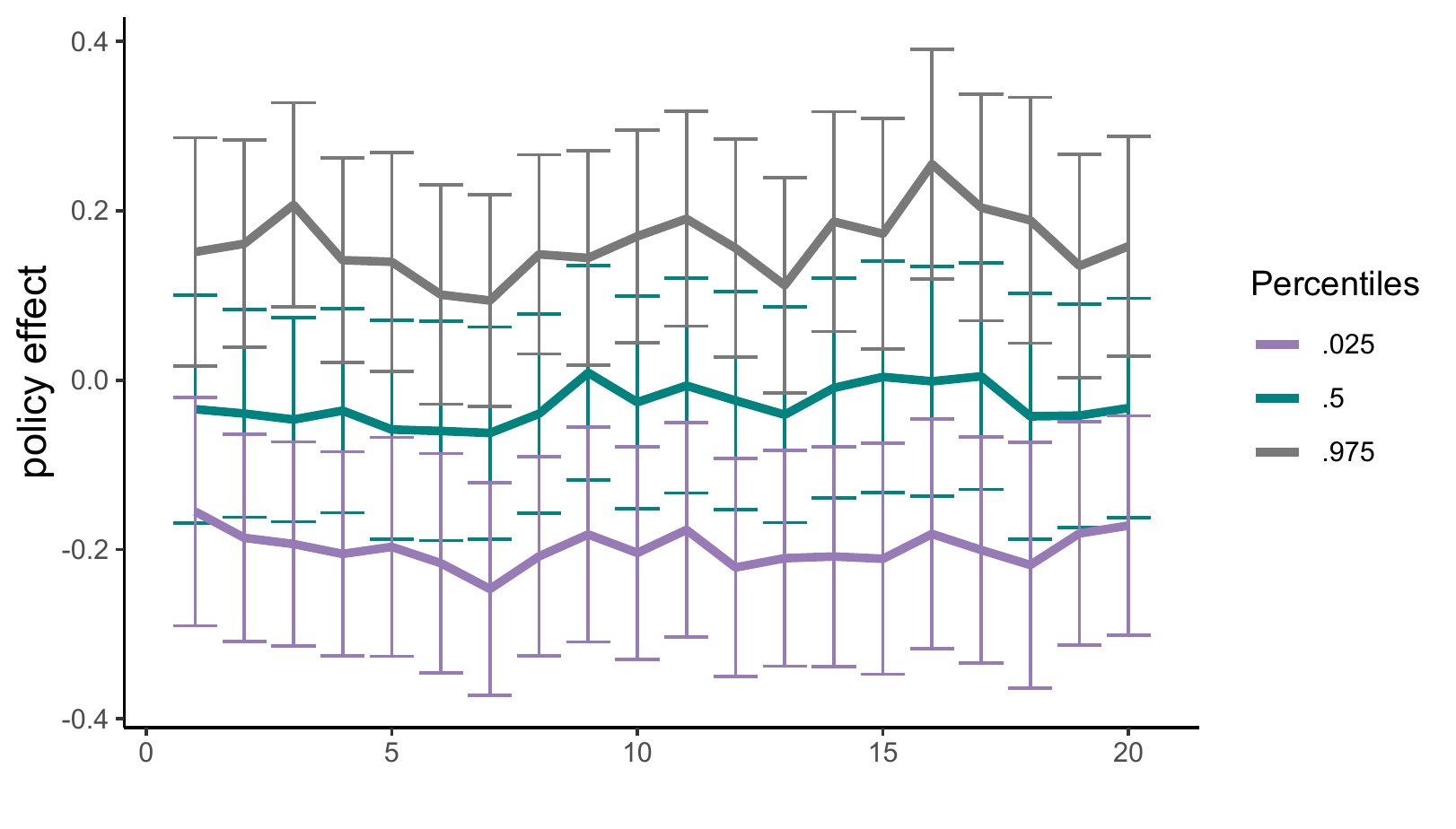}} \quad 
	\caption{Policy effect heterogeneity based on the classical causal forest algorithm (\citealp{wager2018estimation}). The x-axis represents the vigintiles of the corresponding component, while the y-axis measures the "Reemployment Bonus" effect on the logarithm of unemployment duration in weeks. The plot shows the error bars with the color corresponding to a given percentile of the policy effect. The number of trees equals 1000. }\label{fig_group_heterogeneity_cf}
\end{figure}

\end{document}